ORIGINAL ARTICLE

# A step towards the integration of machine learning and classic model-based survey methods

**Tomasz Żądło**[1] · **Adam Chwila**[1]

## Abstract

The usage of machine learning methods in traditional surveys including official statistics, is still very limited. Therefore, we propose a predictor supported by these algorithms, which can be used to predict any population or subpopulation characteristics. Machine learning methods have already been shown to be very powerful in identifying and modelling complex and nonlinear relationships between the variables, which means they have very good properties in case of strong departures from the classic assumptions. Therefore, we analyse the performance of our proposal under a different set-up, which, in our opinion, is of greater importance in real-life surveys. We study only small departures from the assumed model to show that our proposal is a good alternative, even in comparison with optimal methods under the model. Moreover, we propose the method of the ex ante accuracy estimation of machine learning predictors, giving the possibility of the accuracy comparison with classic methods. The solution to this problem is indicated in the literature as one of the key issues in integrating these approaches. The simulation studies are based on a real, longitudinal dataset, where the prediction of subpopulation characteristics is considered.

## 1 Introduction

The model-based approach in survey sampling and small area estimation can be used to make an inference on population and subpopulation characteristics, including, for example, linear combinations, such as the mean, or more complex functions, such as quantiles. The inference can be based on random and non-random samples, including longitudinal surveys, web surveys, and integrated data sets. This approach requires making assumptions about the population distribution of the variable under study, called—in survey sampling—the superpopulation model, see e.g. [1], or shortly the model.

In the model-based approach, different classes of optimal predictors are considered. Suppose that the prediction problem of any linear combination of the variable of interest is analysed. In that case, the Best Linear Unbiased Predictors (BLUPs) and their estimated versions, the Empirical Best Linear Unbiased Predictors (EBLUPs), can be used. BLUPs, considered for example in [2, 3], are predictors which minimise the Mean Squared Error (MSE) in the class of unbiased predictors. The difference between the MSE of the EBLUP and the MSE of the BLUP, resulting from the estimation of model parameters, is usually very small. For example, in simulation studies conducted in [4] based on a dataset from the Polish agriculture census, the obtained MSEs of EBLUPs are higher only by 0.03%–1.1% compared with the MSEs of BLUPs. Hence, looking for more accurate predictors than EBLUPs in the class of unbiased predictors under the correctly specified model is not purposeful. The overview of various EBLUPs modifications, including Spatial EBLUPs (see [5, 6]) and Geographically Weighted EBLUP [7], is presented in the third chapter of [8].

If the objective is to predict any function of the variable of interest by minimising the MSE, the Best Predictor (BP) can be considered as studied in [9]. Hence, the BP is a very useful predictor, but it requires strong distributional assumptions to be met, unlike the EBLUP. Its estimated version,

✉ Tomasz Żądło
tomasz.zadlo@uekat.pl

Adam Chwila
achwila@gmail.com

[1] Department of Statistics, Econometrics and Mathematics, University of Economics in Katowice, 50, 1 Maja Street, Katowice 40-287, Poland

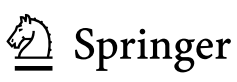

obtained by replacing parameters with estimators, is called the Empirical Best Predictor (EBP). A different approach to the estimation of model parameters in the case of the BP leading to the Observed BP is suggested in [10] and further developed in [11] to incorporate the model selection process, resulting in the Observed Best Selective Predictor (OBSP). In contrast to the differences between the MSEs of EBLUPs and BLUPs, the difference between the MSE of EBP and the MSE of BP can be substantial. Simulation studies presented in [12] based on real data from the U.S. Census Bureau show that the MSEs of EBPs under correctly specified Linear Mixed Model (LMM) are higher by 0.3%–188.2% comparing with the MSEs of BPs. Therefore, looking for predictors more accurately than the EBP is purposeful. It is shown in [12] that in many cases, the PLUG-IN predictors under the LMM are more accurate compared with the EBPs. Further research on the PLUG-IN predictors, which allow predicting any function of the variable of interest, even the distribution function—see, e.g. [13], and do not require strong distributional assumptions, should be conducted. This class of predictors, with the additional robustness property on model misspecification, will be considered.

Machine learning methods offer several advantages over linear models, especially when the deviations from the assumptions of linear models are meaningful. They can model complex nonlinear relationships, automatically capture interaction effects between variables and handle outlier observations, whereas in classic linear models these would need to be manually specified. The advantages of the methods over the linear model in the case of different prediction problems are discussed in [14–16]. In the paper, our considerations will be based on gradient boosting regression trees, which are generally more accurate and train faster comparing with other methods. Moreover, they natively handle categorical variables and outliers without requiring extensive data preprocessing [17]. Since the performance of the methods under strong and moderate departures from classic assumptions is known, it is essential to analyse their performance under different setups, considering small deviations from the assumed model. This is crucial for practitioners who are usually conscious of strong or moderate departures from the classic assumptions but are interested in the performance of the methods they employ when deviations from the model are difficult to identify.

It is important to note that the evaluation of machine learning methods, as discussed in [18, p. 29–30] and [19, pp. 241–249] differs from the accuracy assessment in the model-based approach in survey sampling. In machine learning the aim is usually the prediction of one or more realizations of random variables for future periods or for unobserved population elements. However, the prediction accuracy assessment is based on the differences between predicted and observed real values of the variable, usually via k-fold cross validation, which can be called the ex post prediction accuracy assessment. In model-based approach in survey sampling a function of the random variables (e.g. population or subpopulation mean or median) is predicted. The accuracy assessment is based on difference between random variables (not their realizations as in machine learning): the function of the random variables and its predictor. Hence, the ex ante prediction accuracy is estimated, where *ex ante* does not only mean *future-oriented*, but it is more general, and it refers to non-observed variables in the out-of-sample set. Therefore, integration of the machine learning and model-based methods used in survey sampling requires the ability to assess the prediction accuracy of machine learning methods following the approach used in survey sampling.

Summing up, there are the following aims of the paper:

- the proposal of a PLUG-IN predictor of any population or subpopulation characteristic based on gradient boosting regression trees, which can be used for longitudinal data,
- the comparison of properties of the proposed PLUG-IN predictor with classic methods, including optimal predictors, through a Monte Carlo simulation study under small departures from the assumed model,
- proposals of estimators (parametric, residual and double bootstrap) of ex ante prediction-accuracy measures based on quantiles of absolute prediction errors of the proposed predictor under the model of interest,
- the Monte Carlo comparative simulation studies of the properties of ex ante RMSE estimators and the proposed quantile-based accuracy measures estimators of the proposed predictor and other predictors.

## 2 Survey sampling and machine-learning

In the previous section, we introduced the paper's aims against the background of classic model-based survey methods. In this section, we present a review of the literature on the use of machine learning methods in survey sampling.

According to machine learning methods have become popular among academics and policymakers after the publication of [21] who demonstrated that obtaining socioeconomic information from high-resolution daytime satellite images can help in precise estimation of poverty and wealth. In 2018, a survey on the use of machine learning in official statistics was conducted by the Federal Statistical Office of Germany [22] among 33 national statistical offices, Eurostat, and the OECD. Of those surveyed, 21 institutions reported running projects using machine learning techniques, primarily for data collection and preparation. The usage of the machine learning methods for the inference on the population or subpopulation characteristics, as well as the accuracy

assessment—considered in this paper—was not reported. In the same year, researchers from the UK and Germany proposed a general framework for producing small area statistics [23]. Similar problems were also discussed in [24]. In both publications, the Authors note the relatively low use of machine learning methods in survey sampling compared to other applications. As the critical issue, they define the problem considered in our paper—the ability to obtain correct uncertainty estimates for the considered estimators or predictors.

According to [25], *currently, a paradigm shift from a data modelling culture to an algorithmic modelling culture, as envisioned in* [26], *is taking place in the field of official statistics* [24]. The Authors of the report [27] identified the integration challenges as the critical step in the further development of machine learning methods within the official statistics. The issues considered in this paper can help solve this problem. The proposed procedure for the estimation of the prediction accuracy of new machine learning methods, as it is understood in survey sampling, is absolutely essential in practice. It is a crucial step for the implementation of new predictors in the production of statistics, including the acceptance of the methods by data users. This is because it allows for comparisons with standard methods already used in real-life surveys.

The use of machine learning for small area estimation has generated significant interest in recent literature. Some authors have employed a fully machine learning-based approach for prediction in subpopulations, using data at the subpopulation level (see [14, 28, 29]). This approach differs from traditional small area estimation methods as it does not consider the estimation of prediction accuracy following survey sampling methodology. The approach similar to our proposal is explored in [30], who introduces a new machine learning model: a random forest with the addition of random effects, along with an algorithm to fit the model. Additionally, they introduce a predictor for the small area mean, which can be considered a generalisation of the EBLUP. Finally, they also propose a method similar to the residual bootstrap, for estimating the prediction accuracy under the suggested model. In contrast, our approach allows to predict not only the subpopulation mean but any function of the population (or subpopulation) vector of the variable of interest. We can also compare the prediction accuracy of our method with competitors based on any specified model. Moreover, for predicting subpopulation mean in nonsampled areas, our predictor simplifies to the one proposed in [30, p. 1872].

The use of machine learning methods for small area estimation or survey sampling can be challenging due to issues such as overfitting, small sample sizes, or imbalanced datasets. However, researchers can take control of critical modelling aspects to apply machine learning techniques effectively in these situations. By selecting appropriate modelling techniques based on the characteristics of the dataset (refer to [31] for a comparison of machine learning methods regarding complexity and interpretability), they can optimise their performance. Additionally, the complexity of the model architecture can be controlled through hyperparameters, such as the maximum depth of trees in a random forest or the number of layers in neural networks. In cases with limited sample sizes, researchers can tailor the k-fold cross-validation technique to meet their specific needs [32]. This control over the modelling setup can be crucial in nonstandard situations. Despite these challenges, machine learning techniques are flexible and can adapt well to various datasets, with optimal model architectures evaluated using an independent testing subset. For official statistics, there are other important issues beyond evaluation metrics, including interpretability, model stability, and robustness under different sampling methods. In April 2024, Eurostat launched a four-year project titled "Artificial Intelligence and Machine Learning for Official Statistics". This initiative involves 16 countries and aims to develop a platform for artificial intelligence and machine learning in the realm of official statistics, alongside providing tailored guidance and assistance in employing these solutions within appropriate methodological and implementation frameworks [33].

The application of machine learning methods for small area estimation or survey sampling may be challenging due to overfitting, small sample sizes or imbalanced datasets. Nevertheless, researchers can control crucial aspects of modelling to make use of machine learning techniques even in such cases. The modelling techniques can be chosen adequately for the dataset characteristics (see [31] for a comparison of machine learning methods in terms of complexity and interpretability). Moreover, the complexity of model architecture can be controlled via model hyperparameters such as the maximum depth of trees in random forest, the number of layers of neural networks etc. In the case of a limited sample size, researchers can also adjust the applied k-fold cross-validation technique to their needs [32]. The control over the modelling setup may be essential in case of nonstandard situations. Nevertheless, machine learning techniques are flexible and adapt well to the dataset, while the best model architecture is evaluated on an independent testing subset. For official statistics, there are also other important aspects besides evaluation metrics, such as interpretability, model stability, or robustness under different sampling methods. In April 2024 Eurostat launched a 4-year Project "Artificial Intelligence and Machine Learning for Official Statistics", with involvement of 16 countries to develop a platform for artificial intelligence and machine learning in official statistics, alongside tailored guidance and assistance in deploying artificial intelligence and machine learning solutions within adequate methodological and implementation frameworks [33].

In our studies, gradient boosting regression trees have been applied. However, selecting the most suitable machine learning method for an analysis can be challenging. Utilising gradient boosting regression trees instead of linear mixed models may be particularly beneficial when dealing with outliers in the data. They require less data preparation since they perform well even in the presence of collinearity among variables. Additionally, these trees can identify non-linear relationships, whereas linear mixed models often necessitate data transformations. On the other hand, linear mixed models can incorporate not only plug-in predictors but also Empirical Best Predictors (e.g., [9]). Furthermore, they are generally less demanding on computing power, making them an advantageous choice for simulation studies.

Gradient boosting regression trees often outperform other methods in comparative studies. Research on Kaggle competitions shows that more complex models generally yield better results than simpler ones, with top performance frequently achieved by gradient boosting regression trees and neural networks, including applications involving longitudinal data [34]. Additionally, in small area estimation, gradient boosting regression trees have been found to surpass neural networks, as demonstrated in a study on municipal waste generation in cities [35]. These models also facilitate data interpretation; for instance, Shapley values can be utilized for estimating poverty indicators through social media data [36]. There is even an R package specifically designed for interpreting gradient boosting regression trees with Shapley values [37]. Moreover, if the problem under study requires the application of random effects, the mixed-effects random forest—a method explored for various studies in small area estimation—can be used (see [30, 38, 39]).

In [40] a model-assisted (not model-based) approach is presented that can be treated as an alternative solution but only to estimate the population total. The proposed estimators are generalisations of the generalised regression (GREG) estimator considered in [41]. The Authors replace parametric model-based fitted values in the formula of GREG by fitted values obtained through any parametric or nonparametric procedure. Their approach can be easily extended to apply the machine learning for the estimation of the subpopulation total as well. In [42] in chapters 2.4.2, 2.4.3 and 2.5 it is presented how to modify GREG of the population total to obtain three calibration estimators of the subpopulation totals: with population-specific auxiliary information (given there by equation (2.4.8)), domain-specific auxiliary information (given by (2.4.11)) and modified GREG (given by (2.5.1)). Their formulae can be written as functions of fitted values of a linear model. Replacing them by values fitted by any model (including machine-learning algorithms), the appropriate generalisation will be obtained. What is more, [40] derive the design-variance estimator of their estimator and study its properties in the design-based simulation analyses.

The next two sections will introduce the proposed methodology. In Sect. 3, we will propose a machine learning-based PLUG-IN predictor of any function of the population vector of the variable of interest. This predictor can be used both for cross-sectional and longitudinal surveys. In Sect. 4 we will present our proposal for ex ante accuracy estimation procedure of the predictor.

## 3 Models and predictors

Let the random variable of interest for the $t$ period ($t = 1, 2, \dots, M$) and the $i$th population element ($i = 1, 2, \dots, N_t$), be denoted by $Y_{it}$, where $M$ and $N_t$ are the number of time periods (possibly including the considered future periods) and the population size in the $t$th period, respectively. Let $N_{(L)} = \sum_{t=1}^{M} N_t$. In a special case, when the population does not change in $M$ considered periods ($\forall t\ N_t = N$), then $N_{(L)} = N \times M$. When only cross-sectional population data are considered ($M = 1$), we obtain $N_{(L)} = N$. We assume that longitudinal sample data are available, where the number of observed cases will be denoted by $n_{(L)}$.

Let the population vector of random variables of interest $Y_{it}$, where $t = 1, 2, \dots, M$ and $i = 1, 2, \dots, N_t$, of size $N_{(L)} \times 1$ be denoted by $\mathbf{Y}$. Let the fixed (non-random) matrix of auxiliary variables of size $N_{(L)} \times p$ be denoted by $\mathbf{X}$.

Let us assume that

$$\begin{cases} \mathbf{Y} = m(\mathbf{X}) + \boldsymbol{\xi} \\ E(\boldsymbol{\xi}) = \mathbf{0} \\ Var(\boldsymbol{\xi}) = \mathbf{V} \end{cases}, \tag{1}$$

where $m$ is some fixed but unknown function of auxiliary variables, $\boldsymbol{\xi}$ is a random term with $\mathbf{0}$ mean and unknown variance-covariance matrix $\mathbf{V}$. What is important, the formula (1) covers many models. Let us consider two special cases. Firstly, in machine-learning procedures, see [19, p. 28], usually (1) is considered, where the independence of elements of $\boldsymbol{\xi}$ is additionally assumed. Secondly, the General Linear Mixed Model, which special cases will be analyzed in Sect. 5.1, can also be written as (1). It is given by (e.g. [42], p. 98):

$$\begin{cases} \mathbf{Y} = \mathbf{X}\boldsymbol{\beta} + \mathbf{Z}\mathbf{v} + \mathbf{e} \\ E(\mathbf{e}) = \mathbf{0}, E(\mathbf{v}) = \mathbf{0} \\ Var(\mathbf{e}) = \mathbf{R}(\boldsymbol{\delta}), Var(\mathbf{v}) = \mathbf{G}(\boldsymbol{\delta}) \end{cases}, \tag{2}$$

where $\boldsymbol{\beta}$ is a vector of fixed effects of size $p \times 1$ and $\boldsymbol{\delta}$ is a vector of parameters called the variance components. The random part of the model is described by: a known matrix $\mathbf{Z}$ of size $N_{(L)} \times h$, a vector $\mathbf{v}$ of random effects of size $h \times 1$ and a vector $\mathbf{e}$ of random components of size $N_{(L)} \times 1$, where $\mathbf{e}$ and $\mathbf{v}$ are assumed to be independent. Hence, defining in (1):

- $\boldsymbol{\xi}$ as $\mathbf{Zv} + \mathbf{e}$ (see the random term in (2)),
- and $m(\mathbf{X})$ as $\mathbf{X}\boldsymbol{\beta}$ (see the fixed term in (2))

shows that (2) is a special case of (1), where $\mathbf{V}$ in (1), using the notation used in (2), is given by: $\mathbf{V} = \mathbf{ZGZ}^T + \mathbf{R}$.

Let us assume, without the loss of generality, that the first $n_{(L)}$ elements of $\mathbf{Y}$ are for the sample elements. Then, we can decompose the random vector $\mathbf{Y}$ into the observed and non-observed subvectors: $\mathbf{Y} = [\mathbf{Y}_s^T\ \mathbf{Y}_r^T]^T$, where $\mathbf{Y}_s$ and $\mathbf{Y}_r$ are of sizes $n_{(L)} \times 1$ and $(N_{(L)} - n_{(L)}) \times 1$, respectively. Similarly, we can decompose matrices $\mathbf{X}$ and $\mathbf{Z}$ into: $\mathbf{X} = [\mathbf{X}_s^T\ \mathbf{X}_r^T]^T$ and $\mathbf{Z} = [\mathbf{Z}_s^T\ \mathbf{Z}_r^T]^T$, where $\mathbf{X}_s$, $\mathbf{Z}_s$, $\mathbf{X}_r$ and $\mathbf{Z}_r$ are of sizes $n_{(L)} \times p$, $n_{(L)} \times h$, $(N_{(L)} - n_{(L)}) \times p$ and $(N_{(L)} - n_{(L)}) \times h$, respectively.

Let us consider the problem of prediction of any given function of the population vector of the variable of interest $\theta = \theta(\mathbf{Y}) = \theta([\mathbf{Y}_s^T\ \mathbf{Y}_r^T]^T)$. We consider the PLUG-IN predictor, which for a given $\theta$ is defined as:

$$\hat{\theta} = \theta([\mathbf{Y}_s^T\ \hat{m}(\mathbf{X}_r^T)]^T), \tag{3}$$

where $\hat{m}(\mathbf{X}_r)$ is a $(N_{(L)} - n_{(L)}) \times 1$ vector of fitted values, based on any assumed model, for non-observed random variables. The vector construction will be discussed for two special cases: the General Linear Mixed Model and machine learning algorithms in the following paragraphs in this section.

In the case of the General Linear Mixed Model the fitted values of non-observed random variables, denoted in (3) by $\hat{m}(\mathbf{X}_r)$, are defined as follows $\hat{m}_{GLMM}(\mathbf{X}_r) = \mathbf{X}_r\hat{\boldsymbol{\beta}} + \mathbf{Z}_r\hat{\mathbf{v}}$, where $\hat{\boldsymbol{\beta}}$ and $\hat{\mathbf{v}}$ are given by the formulae of the best linear unbiased estimator of $\boldsymbol{\beta}$ and the best linear unbiased predictor of $\mathbf{v}$ (see [42, p. 98] for more details), respectively, where unknown variance components $\boldsymbol{\delta}$ are replaced by their estimates (e.g. Restricted Maximum Likelihood estimates). What is interesting, the formula of $\hat{m}_{GLMM}(\mathbf{X}_r)$ covers not only $\mathbf{X}_r\hat{\boldsymbol{\beta}}$ but also $\mathbf{Z}_r\hat{\mathbf{v}}$ which results from the assumptions (1), where the random part of the model $\boldsymbol{\xi} = \mathbf{Zv} + \mathbf{e}$ can include spatial or temporal correlations. In machine learning, as stated in [18, p. 17], $\hat{m}(.)$ represents an estimate for $m(.)$, usually treated as a *black box* in the sense that the form of $\hat{m}(.)$ is not of primary interest as opposed to goodness-of-fit and prediction accuracy.

Although any machine learning method can be used, we consider gradient-boosting regression trees—one of the most popular algorithms used for regression problems. It is due to its very good prediction results for real data applications (i.e. Kaggle competitions), relatively low computation time (for example, in comparison with neural networks) and the fact that the algorithm does not require additional data preprocessing like other machine learning methods (including data standardization). The algorithm was introduced simultaneously in 1999 by Jerome H. Friedman [43] and four researchers: Llew Mason, Jonathan Baxter, Peter Bartlett and Marcus Frean []. Algorithm details can be found in [19].

## 4 Ex ante prediction accuracy estimation

In our opinion, in order to integrate machine learning and small-area techniques, it is essential to be able to compare their accuracy in real-life surveys. Therefore, it is crucial to follow the survey methodology to make the comparison. In this section, we will introduce prediction accuracy measures, including our proposal, and procedures of their estimation applying the model-based approach in small area estimation and survey sampling.

### 4.1 Prediction accuracy measures

We consider the problem of prediction of any given function of the population vector $\mathbf{Y}$, denoted by $\theta = \theta(\mathbf{Y})$ by any predictor $\hat{\theta}$, including the PLUG-IN predictor given by (3). Our aim is to assess the accuracy of $\hat{\theta}$ under the Linear Mixed Model (LMM) given by (2) with additional assumption of normality of random effects and random components. However, this approach can be used for any model allowing for the generation of the population vector of the variable of interest, including generalized linear mixed models with logistic mixed model as a special case (see [44, 45]). Let the prediction error be defined as $U = \hat{\theta} - \theta$. The prediction Root MSE (RMSE) is given by

$$RMSE(\hat{\theta}) = \sqrt{E(\hat{\theta} - \theta)^2} = \sqrt{E(U^2)}. \tag{4}$$

Because the MSE is the mean of positively skewed squared prediction errors, we will also use the prediction measure called the Quantile of Absolute Prediction Errors (QAPE) introduced and studied in [46, 47], and defined as:

$$\begin{aligned} &QAPE_p(\hat{\theta}) \\ &= \inf\left\{x : P\left(\left|\hat{\theta} - \theta\right| \leq x\right) \geq p\right\} \\ &= \inf\{x : P(|U| \leq x) \geq p\}. \end{aligned} \tag{5}$$

This measure represents the $p$th quantile of the absolute prediction error $|U|$. It indicates that at least $p \times 100\%$ of absolute prediction errors are smaller or equal to $QAPE_p(\hat{\theta})$, and at least $(1 - p) \times 100\%$ of absolute prediction errors are higher or equal to $QAPE_p(\hat{\theta})$. This means that we are not only interested in average prediction errors but also in the fact that high prediction errors can occur with a certain probability (in $QAPE_p(\hat{\theta})$s of higher orders).

### 4.2 Bootstrap algorithms

The prediction accuracy measures (4) and (5) can be estimated using various bootstrap techniques. Bootstrap algorithms offer flexibility, allowing for the estimation of prediction accuracy under any model of any predictor, not just predictors derived under the model used in the bootstrap. We will consider estimators of (4) and (5) using various, considered below, bootstrap algorithms, to determine which bootstrap algorithm should be preferred in practice for the proposed predictor.

The parametric bootstrap procedure is implemented according to [44, 48] and presented in Appendix A. Based on the procedure, in $B$ iterations we obtain $B$ bootstrap realizations of the prediction errors given by:

$$u^{*(b)} = \hat{\theta}^{*(b)} - \theta^{*(b)}. \tag{6}$$

where $b = 1, 2, \ldots, B$, $\theta^{*(b)}$ is the predicted characteristic computed based on the $b$th bootstrapped population vector of the variable of interest, and $\hat{\theta}^{*(b)}$ is its predictor computed based on the $b$th bootstrapped sample vector of the variable of interest.

The parametric bootstrap estimators of (4) and (5) are given respectively by:

$$\widehat{RMSE}(\hat{\theta}) = \left( B^{-1} \sum_{b=1}^{B} u^{*(b)^2} \right)^{0.5} \tag{7}$$

and

$$\widehat{QAPE}_p(\hat{\theta}) = q_p(|u^{*(1)}|, \ldots, |u^{*(b)}|, \ldots, |u^{*(B)}|), \tag{8}$$

where $u^{*(b)}$, for $b = 1, 2, ...B$, are given by (6), $B$ is the number of bootstrap iterations and $q_p(.)$ is the quantile of order $p$.

To estimate the prediction accuracy, the residual bootstrap procedure can also be used. The detailed description of the algorithm, which can be found in [49–51], is discussed in Appendix A. Residual bootstrap RMSE and QAPE estimators are given by (7) and (8), where parametric bootstrap prediction errors are replaced by the residual bootstrap prediction errors.

The double bootstrap algorithm has been proposed to obtain the bias-corrected MSE estimators. This procedure, studied, among others, in [52–54], consists of two levels, where the parametric bootstrap is used at each level (see Appendix B). At the first level, first-level bootstrap prediction errors given by (6) and the parametric bootstrap MSE and QAPE estimators, given by (7) and (8), are computed. Based on the second-level iterations ($c = 1, 2, \ldots, C$), conducted in each $b$th iteration of the first level, second-level bootstrap prediction errors are computed as

$$u^{**(b,c)} = \hat{\theta}^{**(b,c)} - \theta^{**(b,c)}. \tag{9}$$

where $b = 1, 2, \ldots, B$, $c = 1, 2, \ldots, C$, $\theta^{**(b,c)}$ and $\hat{\theta}^{**(b,c)}$ are the predicted characteristic and its predictor, respectively, computed in the $c$th iteration of the second level within the $b$ iteration of the first level. Based on this procedure, various double bootstrap MSE estimators are presented in [52, 53] (see Appendix B).

We also propose three double bootstrap estimators of QAPE. They are based on the following three proposals of corrected squared first-level bootstrap prediction errors, all presented in [53, p. 3310]:

$$u_1^{**(b)^2} = 2u^{*(b)^2} - C^{-1} \sum_{c=1}^{C} u^{**(b,c)^2}, \tag{10}$$

$$u_2^{**(b)^2} = 2u^{*(b)^2} - u^{**(b,c)^2}, \tag{11}$$

$$u_3^{**(b)^2} = u^{*(b)^2} + u^{*(b+1)^2} - u^{**(b,c)^2}. \tag{12}$$

Because they can be negative, we introduce the following modified double bootstrap prediction errors:

$$u_{i\,mod}^{**(b)} = \begin{cases} \sqrt{u_i^{**(b)^2}} & if \;\; u_i^{**(b)^2} \geq 0 \\ u^{*(b)} & if \;\; u_i^{**(b)^2} < 0 \end{cases}, i = 1, ..., 3, \tag{13}$$

where $u^{*(b)}$ is given by (6), $u_i^{**(b)^2}$ for $i = 1, 2, 3$ are given by (10), (11), and (12), respectively. Based on (13), the following three double bootstrap QAPE estimators are proposed:

$$\widehat{QAPE}_p^{dbC}(\hat{\theta}) = q_p(|u_{1\,mod}^{**(1)}|, \ldots, |u_{1\,mod}^{**(b)}|, \ldots, |u_{1\,mod}^{**(B)}|), \tag{14}$$

$$\widehat{QAPE}_p^{db1}(\hat{\theta}) = q_p(|u_{2\,mod}^{**(1)}|, \ldots, |u_{2\,mod}^{**(b)}|, \ldots, |u_{2\,mod}^{**(B)}|), \tag{15}$$

$$\widehat{QAPE}_p^{dbTel}(\hat{\theta}) = q_p(|u_{3\,mod}^{**(1)}|, \ldots, |u_{3\,mod}^{**(b)}|, \ldots, |u_{3\,mod}^{**(B)}|), \tag{16}$$

where $q_p(.)$ is the $p$th quantile, and values of $u_{i\,mod}^{**(b)}$ for $i = 1, \ldots, 3$ are given by (13) for $i = 1, 2, 3$.

## 5 Monte Carlo simulation studies

In this section, two simulation studies will be conducted. In the first one, the prediction accuracy of the proposed machine learning-based predictor and its competitors will be studied under small departures from the assumed model. In our opinion, large deviations from the classic assumptions are of minor interest in practice, because of two implications. Firstly, they are easier to identify, giving the possibility of the correction of the classic model. Secondly, in such cases, the advantages of machine learning methods over traditional methods are well-known, making them the natural choice (see [14–16]). In cases of substantial or complex non-linearity, we can expect that the proposed predictor based on gradient-boosted regression trees will strongly outperform traditional predictors that rely on linear models, even when data transformations are applied. This advantage stems from the specific capabilities of the gradient-boosted regression trees algorithm, which can effectively capture complex nonlinear dependencies. Non-linear relationships between variables often vary at different levels of the variables, and gradient-boosted regression trees are able to capture these variations, unlike other methods that apply a uniform approach across the entire explanatory variable.

The second simulation study aims to determine whether the comparison of the accuracy of the two approaches is reliable. In our opinion, it will be reliable if: (i) the same approach is used to estimate the accuracy measures for all considered predictors (as proposed in the previous section) and (ii) the properties of estimators of accuracy measures for these predictors are similar and acceptable. The simulation scenario for evaluating the performance of estimators of prediction accuracy measures is based on a linear model. The choice of this model is crucial because, under it, both the considered classic predictor and the estimators of its accuracy measures are correctly specified. Consequently, the properties of the accuracy measure estimators associated with the linear predictor can serve as a benchmark for evaluating the properties of the proposed predictor. If the RMSE and QAPE estimators of the proposed predictor exhibit good performance and are similar to those of the traditional predictor, then this can enhance the reliability of comparing the estimated accuracy of these predictors using real data under the assumed model. Furthermore, we can expect that these conclusions can be valid for any correctly specified model, regardless of its complexity. However, if the bootstrap model is misspecified, its effectiveness in assessing accuracy under an unknown correct model becomes limited for both traditional and proposed predictors. In such cases, further studies on the robustness of these estimators should be conducted.

### 5.1 Dataset and assumptions

We consider a population longitudinal dataset for Polish poviats (until 2016 LAU level 1, formerly NUTS 4) in years 2018–2020 which gives $N \times M = 1140$ observations in total, where the number of periods $M = 3$ and the population size in one period is $N = 380$. Data are freely available via the Statistics Poland's Local Data Bank website (https://bdl.stat.gov.pl). The variable of interest is the average price of $1m^2$ of residential premises in a poviat. The following auxiliary variables are also taken into account: total number of flats ($x_1$), average usable floor area of one flat ($x_2$), average usable floor space per one person ($x_3$), flats per 1000 inhabitants ($x_4$), average number of rooms in one flat ($x_5$), average number of people per one flat ($x_6$), flats put into use per 1000 people ($x_7$), average usable floor space of one flat completed ($x_8$), sale—total number of new notarial deeds ($x_9$).

The aim of the analysis is the prediction of the mean and the median of the variable of interest in the last period in the arbitrarily chosen subpopulation based on the sample data. The subpopulation is Dolnoslaskie voivodeship, the first NUTS 2 region according to Statistics Poland's identifier list. The simulation analysis is fully model-based. The balanced panel is considered, where a simple random sample without replacement is drawn once in the first period ($n = 0.2N$), and the same elements are assumed to be in the sample in the upcoming periods.

In the simulation study, we generate population values of the variable of interest based on the four models—the linear model and three nonlinear models. The considered models, which define four simulation scenarios, are as follows:

- the linear mixed model (denoted by `LM`)

$$Y_{idt} = \beta_{1.pop}x_{1idt} + \beta_{2.pop}x_{4idt} + \beta_{3.pop}x_{7idt} + u_d + e_{idt}, \tag{17}$$

  where $u_d \sim N\left(0, \sigma^2_{u.popL}\right)$, $e_{idt} \sim N\left(0, \sigma^2_{e.popL}\right)$ and the values of the parameters $\beta_{1.pop}$, $\beta_{2.pop}$, $\beta_{3.pop}$, $\sigma^2_{u.popL}$ and $\sigma^2_{e.popL}$ are assumed to be equal to the REML estimates based on (17) and the whole population dataset,
- the first nonlinear mixed model (denoted by `NLM1`)

$$\begin{aligned} Y_{idt} = {} & \beta_{1.pop}\log(x_{1idt}) + \beta_{2.pop}\log(x_{4idt}) + \beta_{3.pop}\log(x_{7idt}) \\ & + \beta_{4.pop}\log(x_{1idt})log(x_{4idt}) + \beta_{5.pop}\log(x_{1idt})\log(x_{7idt}) \\ & + \beta_{6.pop}\log(x_{4idt})\log(x_{7idt}) + v_d + \epsilon_{idt}, \end{aligned} \tag{18}$$

  where $v_d \sim N\left(0, \sigma^2_{v.popN}\right)$, $\epsilon_{idt} \sim N\left(0, \sigma^2_{\epsilon.popN}\right)$ and the values of the parameters $\beta_{1.pop}, \beta_{2.pop}, \cdots, \beta_{6.pop}$, $\sigma^2_{v.popL}$ and $\sigma^2_{\epsilon.popL}$ are assumed to be equal to the REML estimates based on (18) and the whole population dataset,

- the second and the third nonlinear mixed model (denoted by `NLM10` and `NLM20`)

$$
\begin{aligned}
Y_{idt} = {} & \beta_{1.pop}\log(x_{1idt}) + \beta_{2.pop}\log(x_{4idt}) + \beta_{3.pop}\log(x_{7idt}) \\
& + \beta_{4.pop}\log(x_{1idt}) log(x_{4idt}) + \beta_{5.pop}\log(x_{1idt})\log(x_{7idt}) \\
& + \beta_{6.pop}\log(x_{4idt})\log(x_{7idt}) + v_d + e_{idt},
\end{aligned}
\tag{19}
$$

where the values of the parameters $\beta_{1.pop}, \beta_{2.pop}, \cdots, \beta_{6.pop}$, $\sigma^2_{v.popL}$ and $\sigma^2_{\epsilon.popL}$ are the same as in (18), $v_d \sim N\left(0, \sigma^2_{v.popN} \times a^{-2}\right)$, $\epsilon_{idt} \sim N\left(0, \sigma^2_{\epsilon.popN} \times a^{-2}\right)$, where $a = 10$ in the case of model `NLM10` and $a = 20$ in the case of model `NLM20`. It means that the only difference between `NLM1` and models `NLM10` and `NLM20` is that in the case of`NLM10` and `NLM20` the standard deviations of random effects and random components are 10 or 20 times smaller. This makes the nonlinear dependencies and interactions in these models stronger than in the case of `NLM1`.

All the models considered belong to the class of mixed models, which are very popular in survey applications due to their ability to handle complex spatio-temporal correlations in longitudinal data [42]. The nonlinear models included in this study not only address the issue of nonlinearity but also account for possible interactions between independent variables, which are commonly observed in real-world surveys. The parameters used to generate simulation data under model `LM` (see (17)) and `NLM1` (see (18)) are set equal to estimates based on population dataset. As a result, the generated realizations will be similar (refer to the top-left and top-right sections of Fig. 1), despite the differing model forms. Thus, even though model `NLM1` (18) is more complex than model `LM` (17), simulation studies based on `NLM1` can be treated as examples of only minor deviations compared to `LM`. In the cases of models `NLM10` and `NLM20` (see bottom-left and bottom-right sections of Fig. 1), the smaller values of variance components result in stronger nonlinear dependencies. While the departures from the `LM` model remain small, they are more pronounced than those in `NLM1`. The last three models fall into the class of nonlinear models with interactions, which are recognized as effective for modelling various types of data, where interactions are common, and linear models struggle to capture complex relationships between dependent and independent variables (e.g., [55–57]). Specifically, these models represent special cases of the translog model, which has demonstrated successful applications across various fields, including economics, healthcare, and agriculture (e.g., [58–60]). However, using these models as alternatives to a linear model permits coverage of only two potential types of misspecification. Future studies could explore more complex scenarios that address issues such as heteroscedasticity of random effects, alternative distributions of random effects and random components, as well as different variable transformations. However, it is important to emphasise that the goal of this study is to analyse the properties of the proposed methods only under small deviations from classic assumptions. This is because we would like to demonstrate that the machine-learning methods can perform better than the traditional predictors even in such cases. Increasing the number of deviations from the classic model will amplify the differences between the misspecified and correct models. Consequently, machine learning-based methods will be expected to outperform traditional methods to an even greater extent, positioning

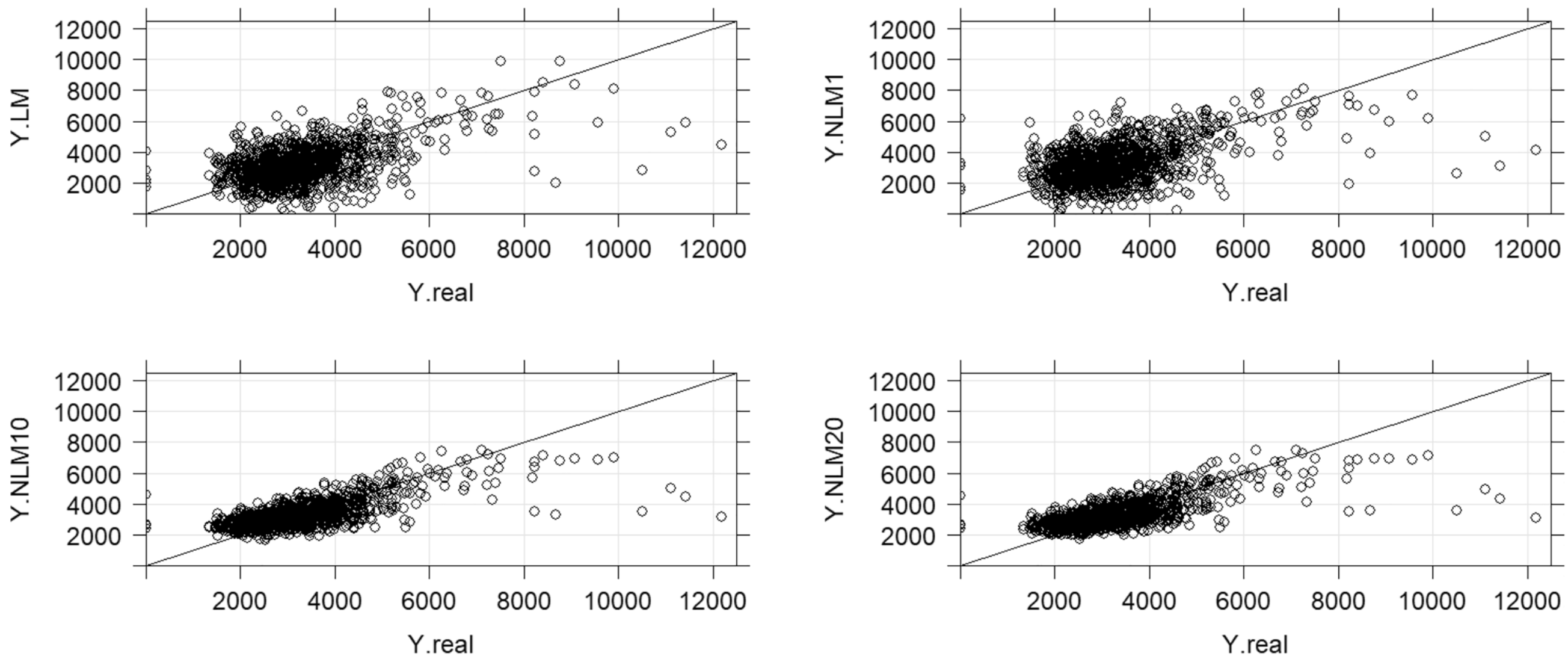

**Fig. 1** One realization of a model versus real values for `LM`, `NLM1`, `NLM10`, and `NLM20`

them as notably more accurate than classic predictors, as discussed in the introduction to Sect. 5.

### 5.2 Simulation study of properties of predictors

In the simulation study of the accuracy of predictors, the number of Monte Carlo iterations is set to be equal $K = 2000$. We study four PLUG-IN predictors (all given by the general formula (3)):

- of the subpopulation mean based on the LMM (17) fitted to the data (in figures denoted by `LMM mean`),
- of the subpopulation mean based on the GB fitted to the data (`GB mean`),
- of the subpopulation median based on the LMM (17) fitted to the data (`LMM median`),
- of the subpopulation median based on the GB fitted to the data (`GB median`).

In the simulation study values of the auxiliary variables in the whole considered dataset are assumed to be known and fixed, while the values of the variable of interest are generated as described in Sect. 5.1. In the case of all predictors used, the same, full set of auxiliary variables $x_1$ - $x_9$, discussed in Sect. 5.1, is treated as the set of potential independent variables. Then, in the case of the considered gradient-boosting PLUG-IN predictor, the process of choosing the auxiliary variables is taken into account in the algorithm. In the case of the PLUG-IN predictor based on the linear mixed model, we use the permutation tests (see [61]) to test their significance. Based on the test procedure, under 0.05 significance level, we can state, that $x_1$, $x_4$ and $x_7$ have a significant influence on the variable of interest and these variables are used to compute the fitted values based on the LMM.

In the case of the PLUG-IN predictor based on the GB tree algorithm, the model fitted values are computed using `xgboost R` library [62]. To fit the model to the data, we consider 7 hyperparameters, which are the standard ones available in this package: maximum number of iterations of gradient boosting algorithm, eta (learning rate), gamma (regularization parameter that controls overfitting), maximum depth of the tree, minimum number of observations in the single tree leave, percentage of auxiliary variables randomly included in the single tree, percentage of observations randomly selected for the single tree. The final set of hyperparameters is selected with the usage of 5-folded cross-validation [19, p. 241]). In our case, the sample in the first period is randomly divided into 5 segments of approximately equal sizes (20% of the sample size in the first period), and the chosen observation remains in the same segment in the rest of the periods throughout the cross-validation process. The hyperparameters selection is performed with the random search algorithm [63], and the final set of hyperparameters is chosen among 1000 sets of random hyperparameters. In the case of the PLUG-IN predictor based on the Linear Mixed Model, the model parameters are estimated using the Restricted Maximum Likelihood Method [64] implemented in the `lme4 R` library [65].

In Fig. 2 we present the following relative measures allowing for the accuracy assessment of the predictors:

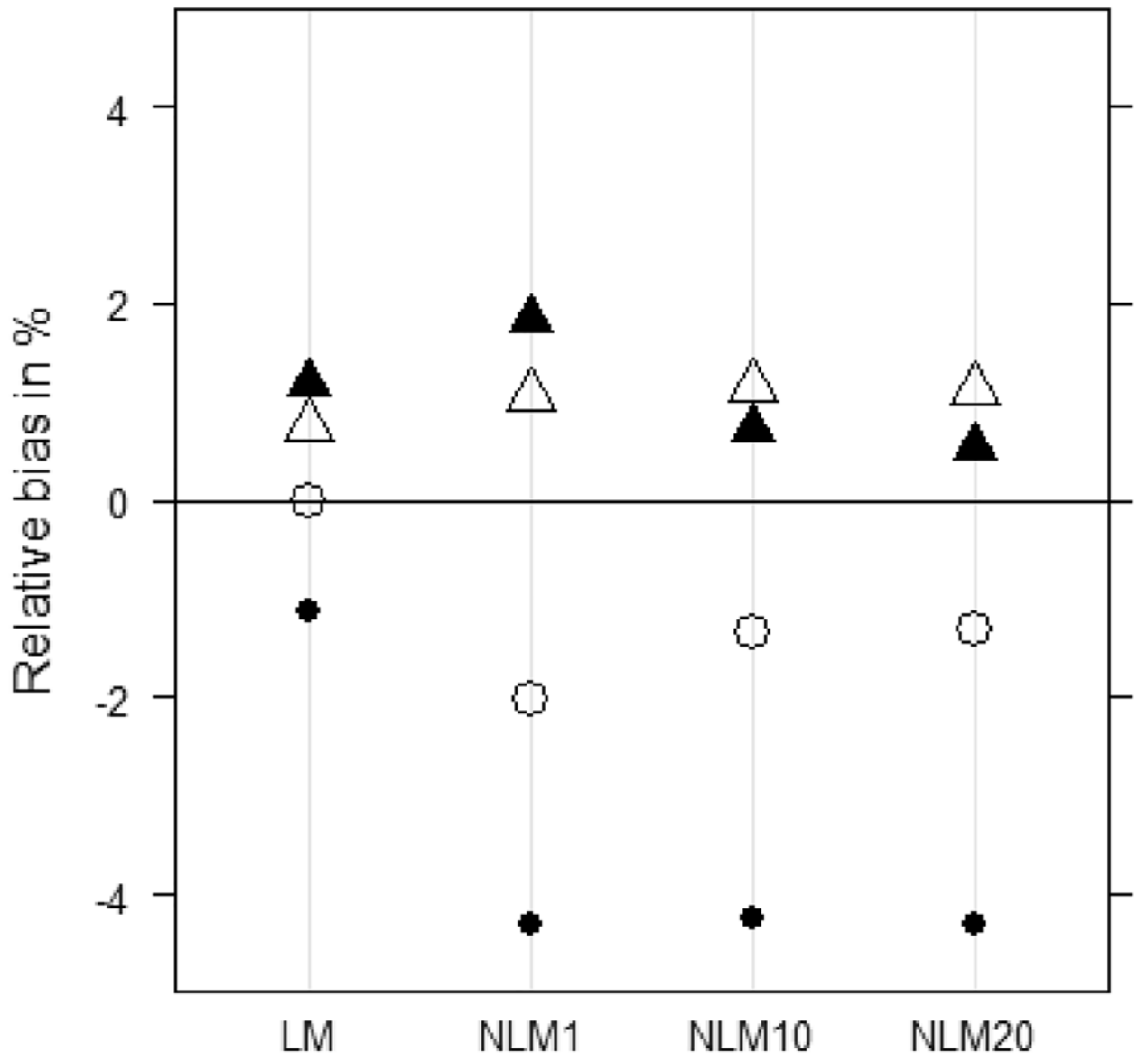

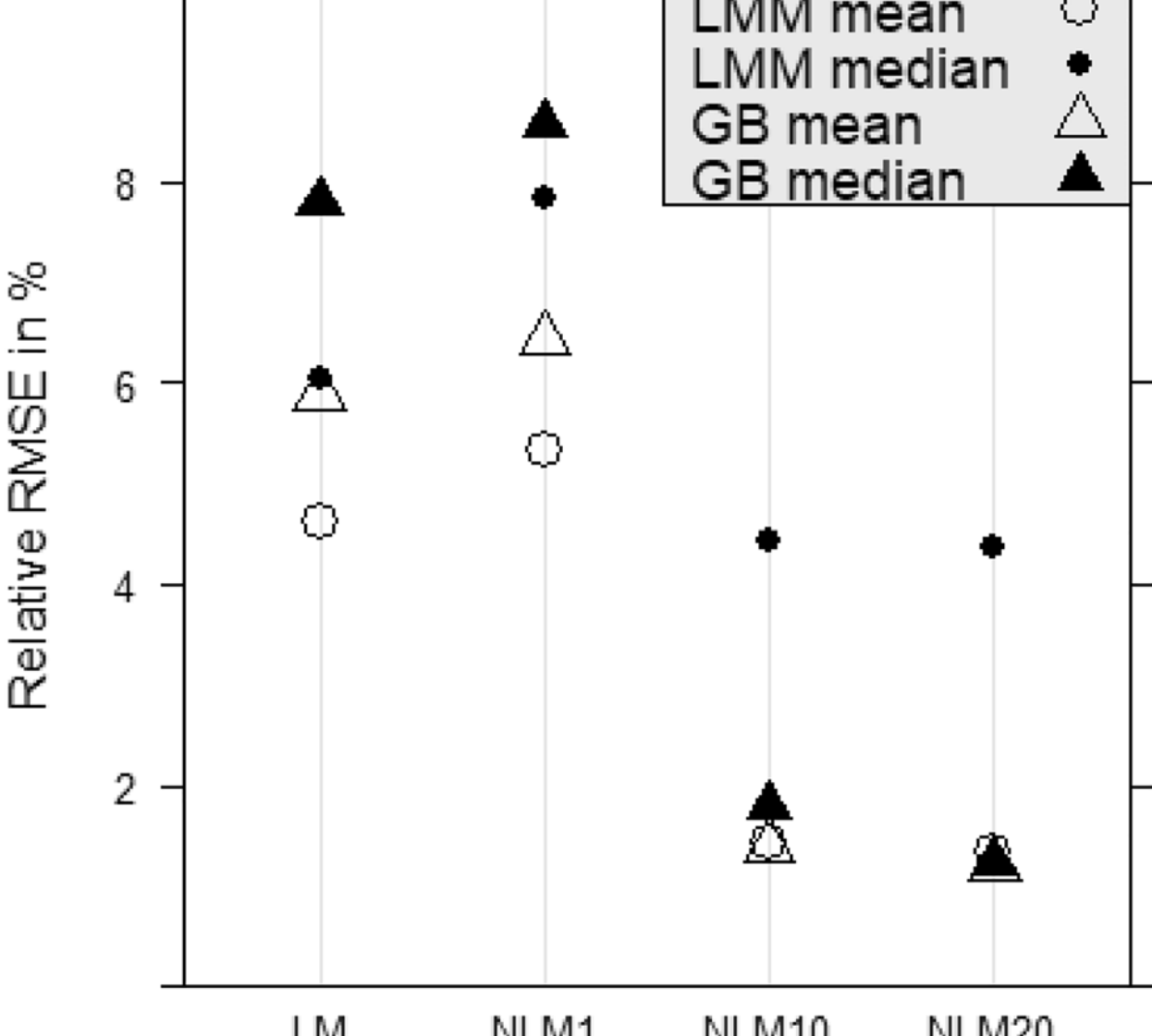

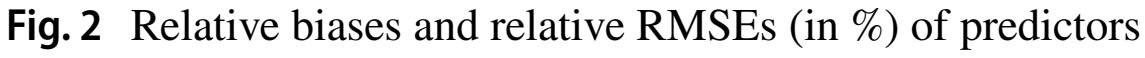

**Fig. 2** Relative biases and relative RMSEs (in %) of predictors

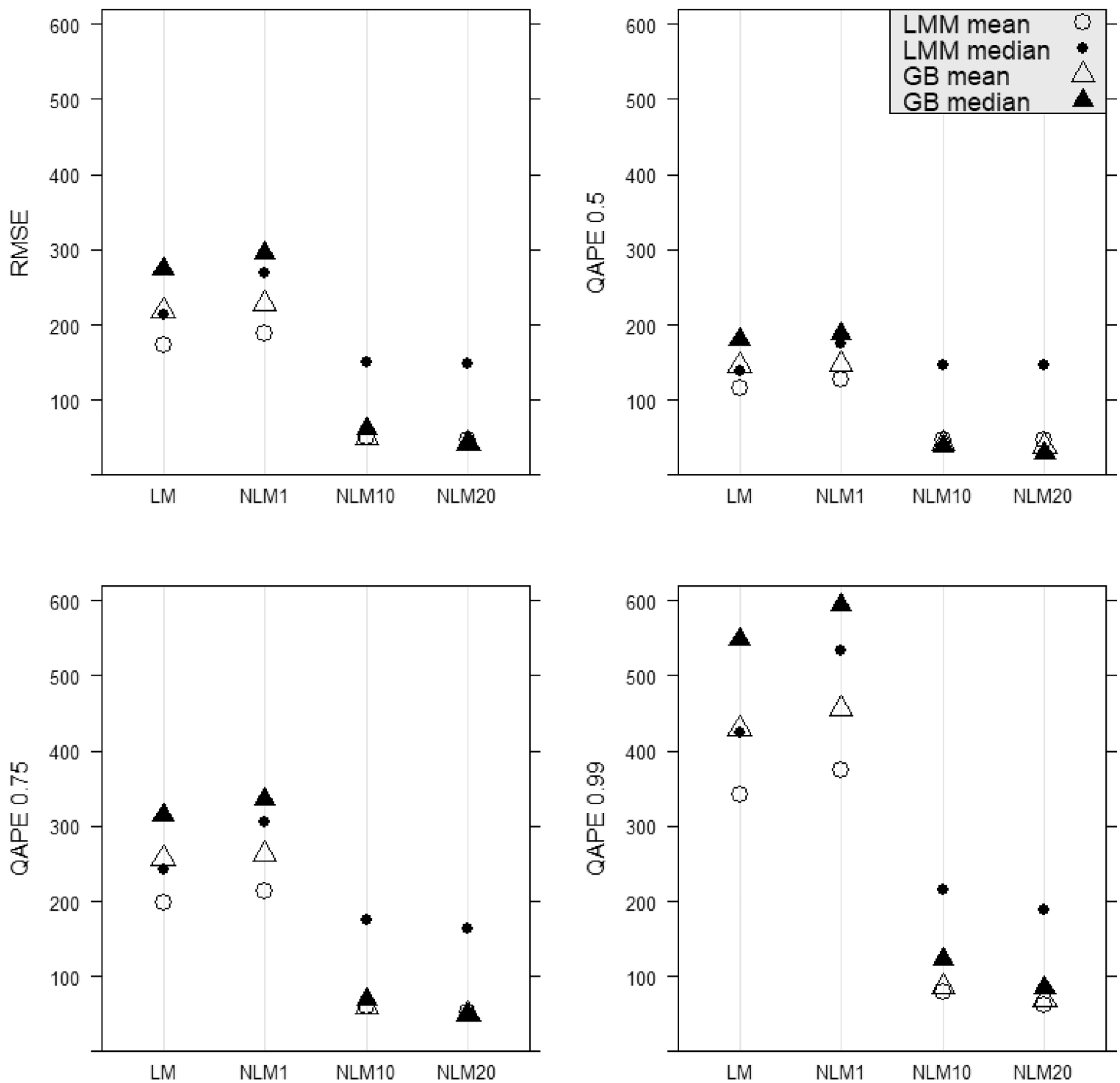

**Fig. 3** RMSEs and QAPEs of predictors

- the relative prediction bias

$$rB^{(sym)}\left(\hat{\theta}\right) = \frac{\frac{1}{K}\sum_{k=1}^{K}(\hat{\theta}^{(k)} - \theta^{(k)})}{\frac{1}{K}\sum_{k=1}^{K}\theta^{(k)}}100, \tag{20}$$

- the prediction relative RMSE (rRMSE)

$$rRMSE^{(sym)}\left(\hat{\theta}\right) = \frac{\sqrt{\frac{1}{K}\sum_{k=1}^{K}\left(\hat{\theta}^{(k)} - \theta^{(k)}\right)^2}}{\frac{1}{K}\sum_{k=1}^{K}\theta^{(k)}}100. \tag{21}$$

In Fig. 3 we present the computed values of absolute prediction measures:

- the RMSE

$$RMSE^{(sym)}\left(\hat{\theta}\right) = \sqrt{\frac{1}{K}\sum_{k=1}^{K}\left(\hat{\theta}^{(k)} - \theta^{(k)}\right)^2}, \tag{22}$$

- the quantile of absolute prediction error (QAPE) of order $p$ introduced in Sect. 4.1 (see (5))

$$QAPE_p(\hat{\theta}) = q_p(|\hat{\theta}^{(1)} - \theta^{(1)}|, \ldots, |\hat{\theta}^{(k)} - \theta^{(k)}|, \ldots, |\hat{\theta}^{(K)} - \theta^{(K)}|), \quad (23)$$

where $q_p(.)$ is the $p$th quantile.

Firstly, let us consider the results under the correctly specified model where the mean is predicted, which means that the values of the variable of interest are generated based on `LM` model and the same model is used to construct `LMM mean` predictor, which is the EBLUP in this case. It means it is the empirical (estimated) version of the optimal predictor, in the sense that the predictor minimizes the prediction MSE in the class of the unbiased predictors. We can see these theoretical properties in our simulation results. The relative simulation biases presented on the left part of Fig. 2 of the `LMM mean` EBLUP predictor (denoted by "∘" symbol) under `LM` are very close to zero, and—see the right part of Fig. 2—the rRMSE of the predictor is smaller, under `LM` model, comparing with another mean predictor—`GB mean` (denoted by "△" symbol). Similarly, in Fig. 3, values of all considered accuracy measures for `LMM mean` predictor are smaller comparing with `GB mean` under `LM` model. To sum up, although `LMM mean` predictor is optimal in the discussed sense, the `GB mean` predictor is only slightly less accurate in this case.

Thirdly, the problem of model misspecification is studied. Under models `NLM1`, `NLM10` and `NLM20`, the absolute biases of the proposed GB-based predictors are smaller than the respective absolute biases of the LMM-based predictors (see the left part of Fig. 2). As shown on the right part of Fig. 2 and in Fig. 3, under model `NLM1`, even though the model is misspecified, the accuracy of LMM-based predictors is still slightly better than the accuracy of GB-based predictors. However, under `NLM10` and `NLM20` the accuracy of GB-based predictor of the mean measured by the RMSE (and rRMSE) and QAPEs of orders 0.5 and 0.75 is better by up to 15% (if measured by QAPE of order 0.5 for `NLM20`), and in the case of QAPE of order 0.99—very similar. For the same models, GB-based predictor of the median is from 1.75 to 5.25 times more accurate comparing with the LMM-based predictor of the median, where the results depend on the accuracy measure.

Summing up, we have shown that the GB-based predictors are very good alternatives to the LMM-based predictors, including optimal predictors. They provide only slightly less accurate results under the correctly specified LMM, and better results even for small departures from the assumed models. Therefore, the next crucial step is to be able to properly compare the accuracy of the new predictor with its competitors based on sample data. Without this step it is not possible to use the proposed predictor appropriately in practice. The proper comparison indicates that the proposed estimators of accuracy measures of the proposed predictor should have very good properties that are similar to the estimators of accuracy measures of the competitive predictors.

### 5.3 Simulation study of properties of accuracy measures estimators

In this simulation study, the properties of RMSE and QAPE estimators are analysed under (17) as motivated in the introduction of Sect. 5. The assumed number of Monte Carlo iterations is $K = 1000$; the number of the parametric, residual and the first level of double bootstrap iterations equals $B = 200$; and the number of the double bootstrap second level iterations is assumed to be $C = 1$. The assumed value of $C$ is set due to the time-consuming computations, but it is shown to be the best choice in the case of the EBP. This conclusion is based on the simulation studies presented by [53] pp. 3315–3316, where several equally efficient bootstrap designs are examined—designs that achieve the same efficiency as the MSE estimators produced for $C = 1$. The authors demonstrated that the number of iterations required to obtain MSE estimators with similar efficiency to those for $C = 1$ is up to 26 times higher. Therefore, the design with $C = 1$ is identified as the least demanding in terms of computational power while providing the same level of efficiency. Although the considered PLUG-IN predictor is similar to the EBP, in further research other values of $C$ can be considered as well, especially if the double bootstrap procedure will occur to be the preferable method based on the results of the conducted simulation analysis.

We study the properties of RMSE estimators based on:

- parametric bootstrap, given by (7) (denoted below by `param`),
- residual bootstrap with and without the correction, given by (7), where parametric bootstrap prediction errors are replaced by appropriate residual bootstrap prediction errors (`rbCor` and `rb`),
- double bootstrap, all considered in [53, p.3310–3311], given in Appendix B by (B6), (B8), (B10) with $C = 1$, (B11) and (B12) (denoted by `db1`, `dbTel`, `db1HM`, `db1EF`, and `dbTelEF`, respectively),

and QAPE estimators based on:

- parametric bootstrap, given by (8) (`param`),
- residual bootstrap with and without the correction, given by (8), where parametric bootstrap prediction errors are replaced by appropriate residual bootstrap prediction errors (`rb` and `rbCor`),
- double bootstrap, given by (15) and (16) (`db1` and `dbTel`).

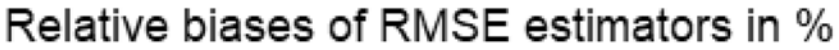

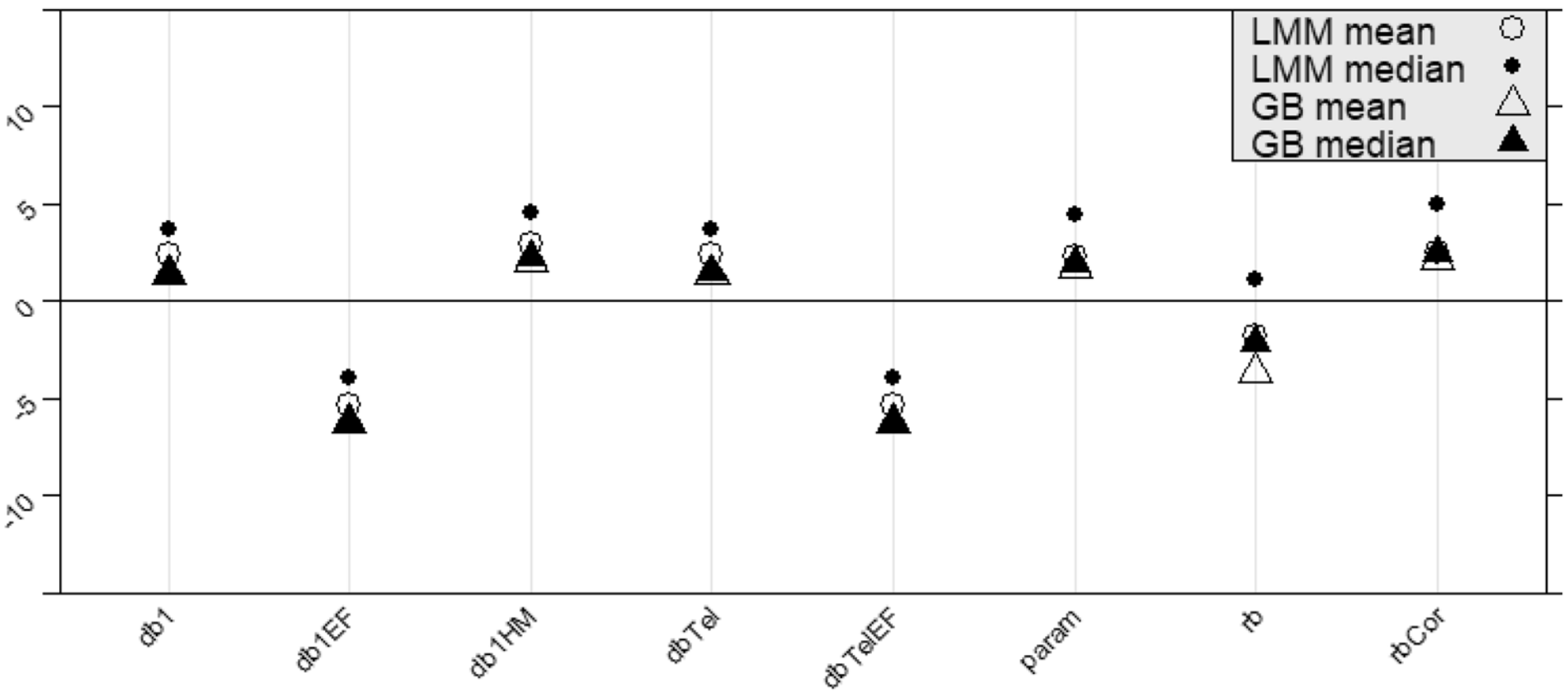

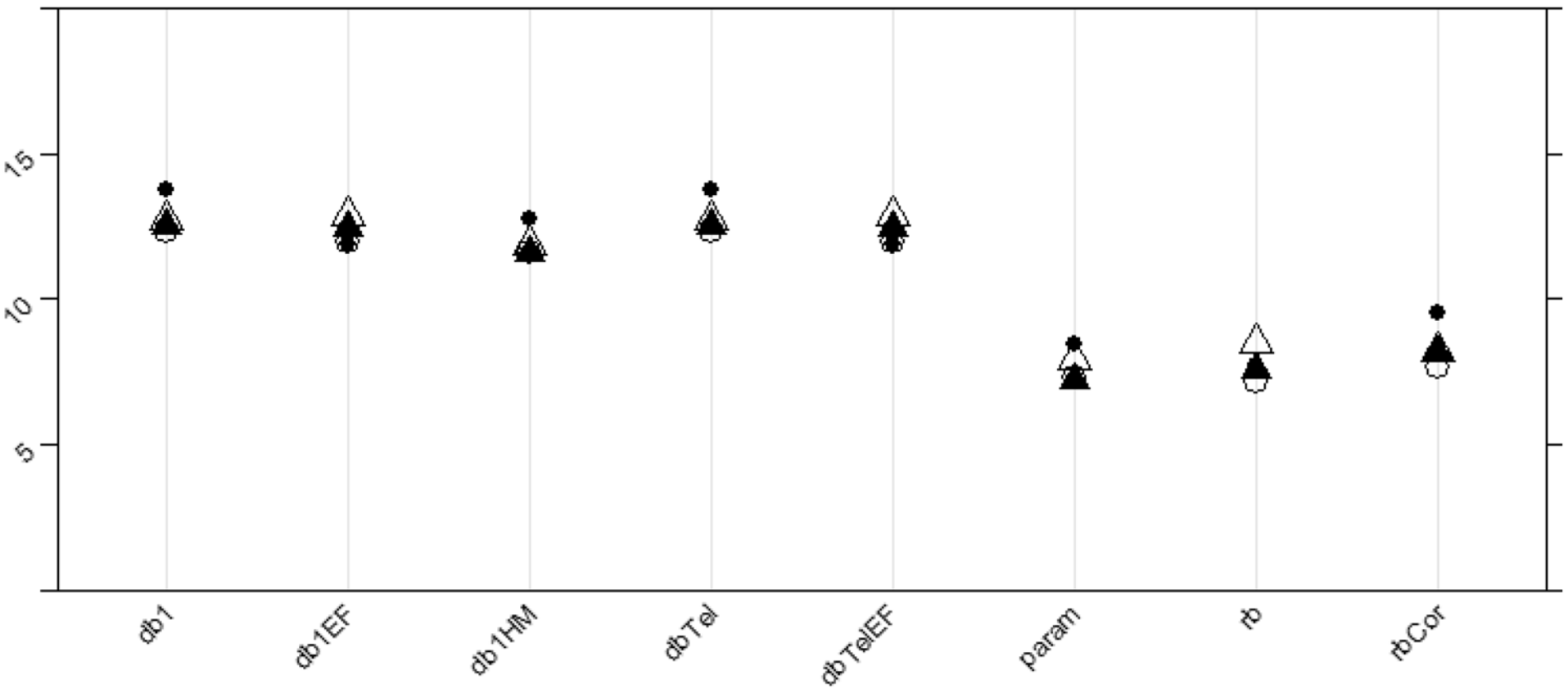

**Fig. 4** Relative biases and relative RMSEs (in %) of RMSE estimators

We present values of their relative biases and relative RMSEs. They are computed based on (20) and (21), respectively, where $\hat{\theta}^{(k)}$ are replaced by the values of the bootstrap RMSE estimators (or bootstrap QAPE estimators) obtained in the $k$th iteration of the simulation study, and $\theta^{(k)}$ are replaced by the RMSE given by the square of (22) (or by the QAPE given by (23)).

Firstly, let us analyse the biases of RMSE estimators presented in the top part of Fig. 4 and of QAPE estimators presented in the top parts of Figs. 5 and 6. For all the considered cases, both for GB-based and LMM-based predictors, the parametric and residual (with and without correction) bootstrap algorithms lead to only slightly biased RMSE and QAPE estimators. The double bootstrap method was originally developed to correct the bias of the MSE estimators for the Empirical Best Predictor. However, for the RMSE estimators of the considered predictors, it produces a similar or larger bias compared to other RMSE estimators based on different bootstrap methods. The situation is even worse for QAPE estimators, as double bootstrap generates higher biases of QAPE estimators than other methods. Secondly, the accuracy of RMSE estimators presented in the bottom part of Fig. 4 and of QAPE estimators presented in the bottom parts of Figs. 5 and 6 is analysed. In this case, the smallest values of relative RMSEs are observed if the parametric and both residual bootstrap algorithms are used—relative RMSEs do not exceed: for RMSE estimators 10%, for QAPE(0.5) estimators 14%, and for QAPE(0.99) estimators 11%. In this comparison, we found that the double bootstrap RMSE and QAPE estimators are notably less accurate than their competitors for the considered predictors. Therefore, we can expect that further research on the properties of the double bootstrap RMSE and QAPE estimators under conditions besides $C = 1$ would be inefficient. In the case of RMSE estimators, they exhibit similar (and acceptable) biases as other methods, indicating that further reduction of bias is unnecessary. For the QAPE double bootstrap estimators, reducing bias is necessary, suggesting that additional studies on the choice of $C$ could be useful. However, reducing bias typically leads to decreased accuracy, which may result in highly inefficient estimators, which currently are even several times less accurate than competitors.

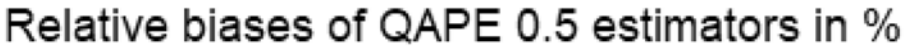

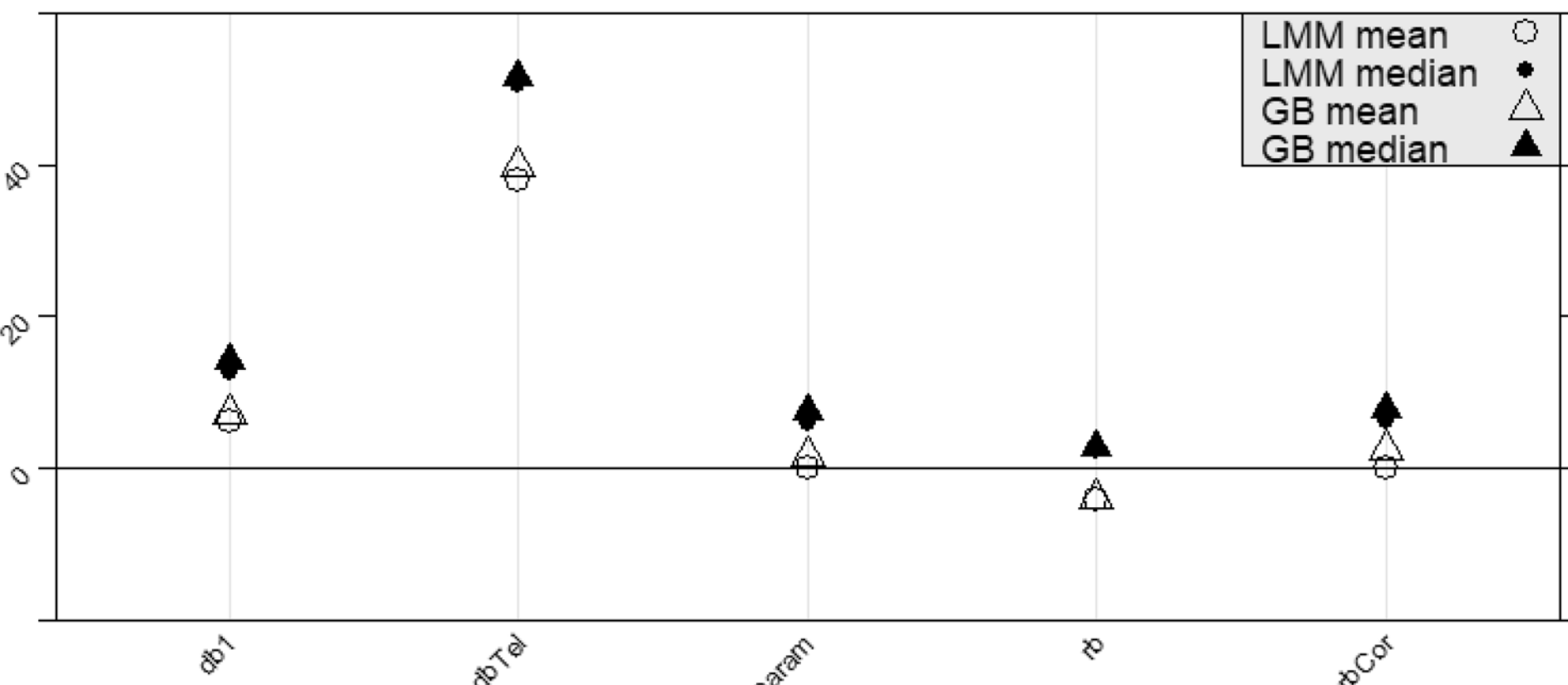

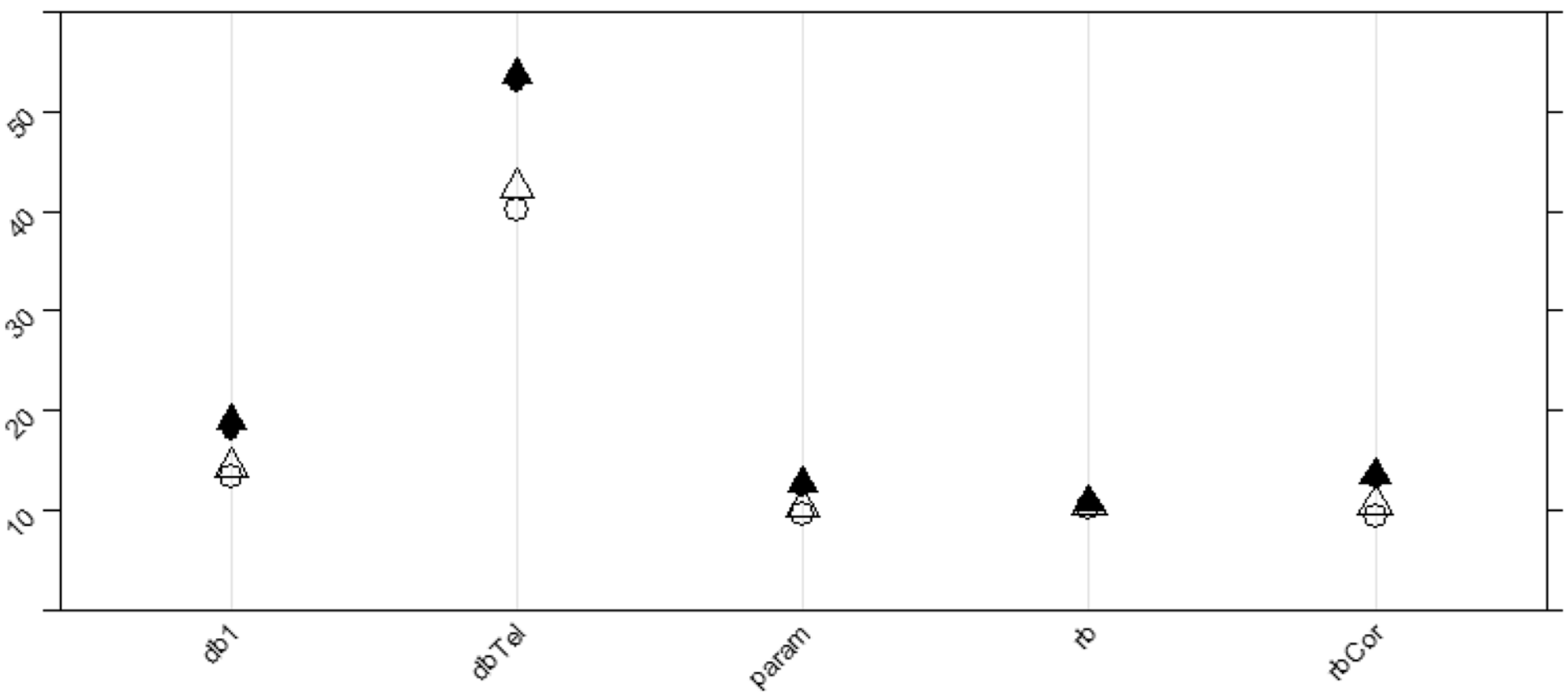

**Fig. 5** Relative biases and relative RMSEs (in %) of QAPE(0.5) estimators

Furthermore, it is important to note that additional studies may actually seem unnecessary, as the results obtained for both the parametric and residual bootstrap methods are shown to be satisfactory.

Summing up, for each of the considered bootstrap algorithms the properties of a certain RMSE or QAPE estimator are very similar irrespective of the predictor (GB-based or LM-based) and the prediction problem (the prediction of the mean or the median). What is more, RMSE and QAPE estimators based on three bootstrap algorithms, namely the parametric and two residual bootstrap methods, have very good properties. It means that using them, we can accurately estimate the accuracy of the proposed predictor and reliably compare it with the estimated accuracy of different predictors. This, in our opinion, paves the way for the possibility of practical use of the proposed predictor in practice, where it is important not only to be able to assess the population or subpopulation characteristics but also to estimate the prediction accuracy and to compare the prediction accuracy estimates.

For better coverage of multiple dimensions of real-world complexity, this study has covered four different scenarios to enhance the joint analysis of the accuracy of the predictors and the performance of the ex ante accuracy measure estimators. However, it is important to note that the conclusions drawn from any Monte Carlo analysis are limited to the dataset and setups considered in the research, regardless of the number of simulation settings examined. Consequently, if different datasets or simulation scenarios are analysed, further simulation studies are required.

## 6 Conclusion

We proposed a predictor of any population or subpopulation characteristic based on gradient-boosting regression trees. We showed that under the LMM the accuracy of the proposed predictor is similar to the accuracy of the PLUG-IN predictor based on the LMM, but it is better regarding even relatively small departures from the linearity.

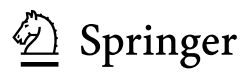

**Fig. 6** Relative biases and relative RMSEs (in %) of QAPE(0.99) estimators

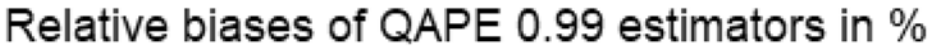

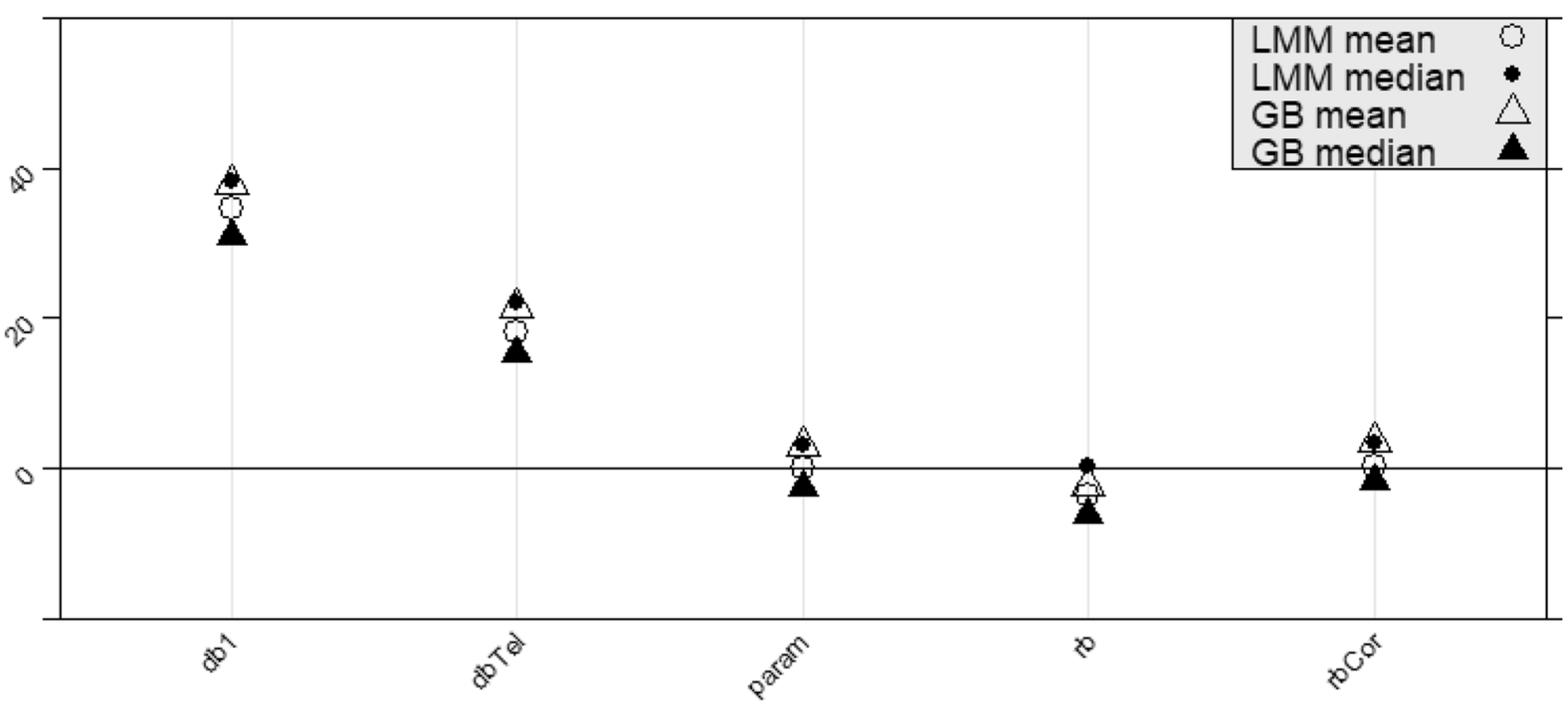

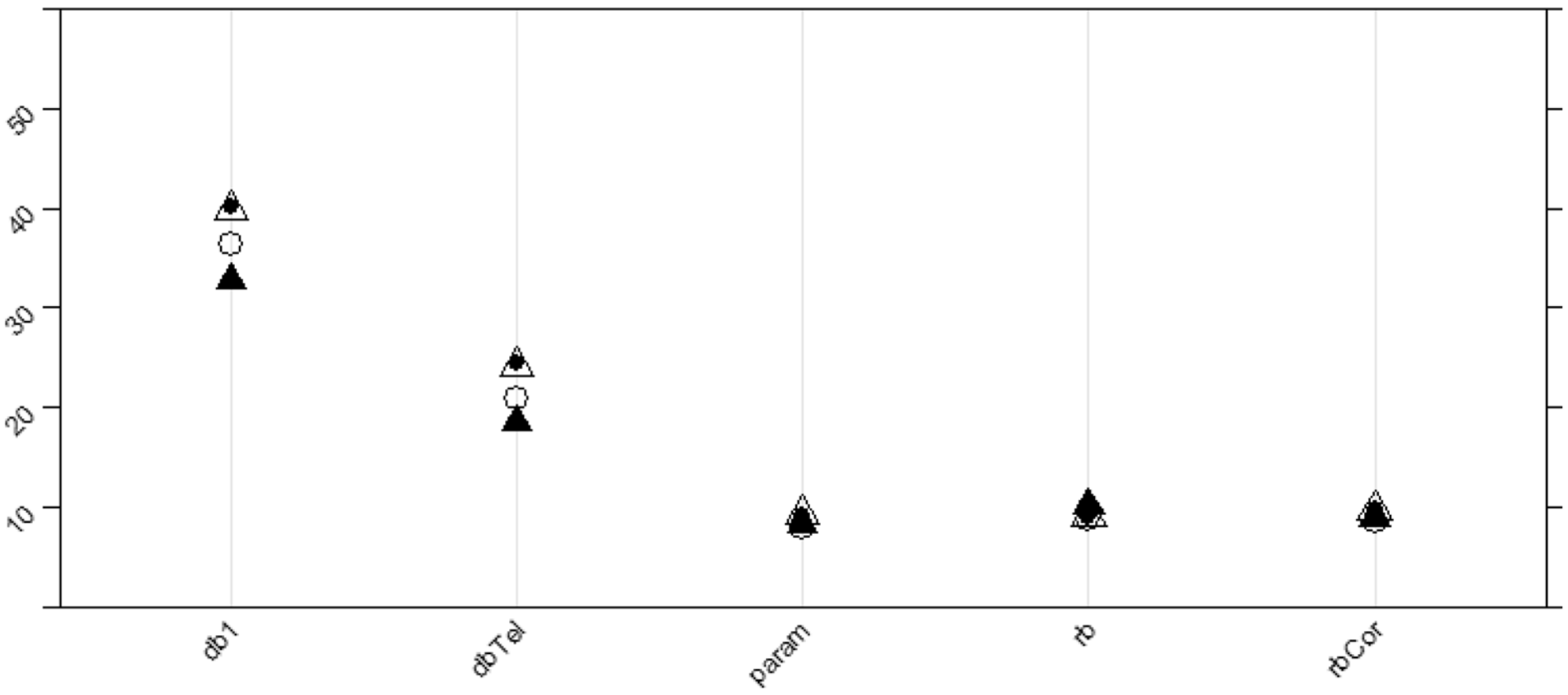

Therefore, we lose little under the correctly specified model, and we can gain a lot (we obtained up to 5.25 times more accurate results) under an even slightly misspecified model.

What is more, the properties of all studied accuracy measures estimators under the LMM are very similar for both predictors. Hence, in practice, we can compare estimators of accuracy measures of the proposed predictor and classic predictors based on the LMM. Finally, we showed very good properties of the parametric and residual bootstrap RMSE and QAPE estimators, which allows us to recommend them for empirical research.

## Parametric and residual bootstrap procedures

The parametric bootstrap procedure is implemented according to [44, 48] and could be described in the following steps.

(a) Based on $n_{(L)}$ sample observations of the dependent and independent variables, model parameters are estimated.

(b) Based on $N_{(L)}$ population observations of the independent variables, a realization of the population vector of the dependent variable of size $N_{(L)} \times 1$ is generated under the assumed model, where parameters are replaced by their estimates (e.g. Restricted Maximum Likelihood estimates) and under normality of random effects and random components.

(c) The population vector of the dependent variable generated in the previous step is decomposed into two subvectors: the first of size $n_{(L)} \times 1$ for the sample observations, and the second of size $(N_{(L)} - n_{(L)}) \times 1$ for non-sampled observations.

(d) Based on the generated population vector of the dependent variable, the bootstrap realization of the predicted characteristic, denoted for the $b$th iteration by $\theta^{*(b)}$, is computed.

(e) The generated sample vector of the dependent variable is used to compute the vector of estimates of model parameters, and based on these vectors, the bootstrap realization of the predictor $\hat{\theta}$, denoted for the $b$th iteration by $\hat{\theta}^{*(b)}$, is computed.

(f) The bootstrap realization of the prediction error is calculated as

$$u^{*(b)} = \hat{\theta}^{*(b)} - \theta^{*(b)}. \tag{A1}$$

(g) Steps (b)-(f) are repeated $B$ times.

A detailed description of the algorithm can be found in [49–51]. To obtain the residual bootstrap procedure, the step (b) in the parametric bootstrap algorithm presented above should be replaced with:

(b) Based $N_{(L)}$ on population observations of the independent variables, estimated fixed effects, and simple random samples with replacement of predicted random effects and estimated random components, a realization of the population vector of the dependent variable of size $N_{(L)} \times 1$ is generated under the assumed model.

If the model covers more than one vector of random effect at the same grouping level, then the predicted values of these effects for the same level are sampled jointly (rows of the matrix formed by these vectors are sampled with replacement). The residual bootstrap algorithm can also be performed with a so-called "correction procedure" [51, p. 132] to improve the properties of the residual bootstrap estimators due to the underdispersion of the uncorrected residual bootstrap distributions.

## Double bootstrap procedure

As presented in [52–54], the double bootstrap procedure consists of two parametric bootstrap levels. For the $b$th iteration of the first level, described in Appendix A, the following second level is conducted.

In the $c$th iteration ($c = 1, 2, ..., C$) of the second level:

(a) Model parameters are estimated Based on $n_{(L)}$ sample observations of the dependent variable generated at the first level and independent variables.

(b) Based on $N_{(L)}$ population observations of the independent variables, a realization of the population vector of the dependent variable of size $N_{(L)} \times 1$ is generated under the assumed model where parameters are replaced by their estimates (obtained in the previous step of the second level bootstrap procedure) and under normality of random effects and random components.

(c) The population vector of the dependent variable generated in the previous step is decomposed into two subvectors: the first of size $n_{(L)} \times 1$ for the sample observations, and the second of size $(N_{(L)} - n_{(L)}) \times 1$ for non-sampled observations.

(d) Based on the generated (at the second level bootstrap procedure) population vector of the dependent variable, the bootstrap realization of the predicted characteristic, denoted by $\theta^{**(b,c)}$, is computed.

(e) The generated (at the second level bootstrap procedure) sample vector of the dependent variable is used to compute the vector of estimates of model parameters, and based on these vectors, the bootstrap realization of the predictor $\hat{\theta}$, denoted by $\hat{\theta}^{**(b,c)}$, is computed.

(f) The second-level bootstrap prediction error is computed as

$$u^{**(b,c)} = \hat{\theta}^{**(b,c)} - \theta^{**(b,c)}. \tag{B2}$$

(g) Steps (b)-(f) are repeated $C$ times.

The following double bootstrap MSE estimators are considered in the literature. The classic double-bootstrap estimator, considered in [52, p. 228] and [53, p. 3310], where the number of second level bootstrap iterations $C \geq 1$, is given by:

$$\widehat{MSE}^{dbC}(\hat{\theta}) = 2\widehat{MSE}^{param} - \widehat{MSE}^{db-2lev} = B^{-1}\sum_{b=1}^{B} u_1^{**(b)^2}, \tag{B3}$$

where

$$\widehat{MSE}^{db-2lev} = B^{-1}C^{-1}\sum_{b=1}^{B}\sum_{c=1}^{C} u^{**(b,c)^2}, \tag{B4}$$

$$u_1^{**(b)^2} = 2u^{*(b)^2} - C^{-1}\sum_{c=1}^{C} u^{**(b,c)^2} \tag{B5}$$

and $u^{*(b)}$ and $u^{**(b,c)}$ are given by (6) and (9), respectively. Its special case proposed in [66] (compare [53, p. 3310]), where $C = 1$, is as follows:

$$\widehat{MSE}^{db1}(\hat{\theta}) = B^{-1}\sum_{b=1}^{B} u_2^{**(b)^2}, \tag{B6}$$

where

$$u_2^{**(b)^2} = 2u^{*(b)^2} - u^{**(b,c)^2}. \tag{B7}$$

In [53, p. 3310] a modification of (B6) called the telescoping bootstrap MSE estimator is proposed. It is given by:

$$\widehat{MSE}^{dbTel}(\hat{\theta}) = B^{-1}\sum_{b=1}^{B} u_3^{**(b)^2}, \tag{B8}$$

where

$$u_3^{**(b)^2} = u^{*(b)^2} + u^{*(b+1)^2} - u^{**(b,c)^2}. \tag{B9}$$

According to formula (12), the number of first-level bootstrap prediction errors to be computed for (B8) is $B + 1$.

Due to observed in simulation studies, possible unacceptable bias corrections included in the above formulae, which can lead even to negative values of MSE estimators, modifications of (B3), (B6) and (B8) are proposed. A modification of (B3), with the number of second level iterations $C \geq 1$, considered in [52, p. 228] is as follows:

$$\widehat{MSE}^{dbCHM}(\hat{\theta}) = \begin{cases} 2\widehat{MSE}^{param} - \widehat{MSE}^{db-2lev} & if \ \widehat{MSE}^{param} \geq \widehat{MSE}^{db-2lev} \\ \widehat{MSE}^{param} exp\left[\frac{\widehat{MSE}^{param} - \widehat{MSE}^{db-2lev}}{\widehat{MSE}^{db-2lev}}\right] & if \ \widehat{MSE}^{param} < \widehat{MSE}^{db-2lev} \end{cases} \tag{B10}$$

In [53] the following modification of (B6) is proposed:

$$\widehat{MSE}^{db1EF}(\hat{\theta}) = \begin{cases} q \times \widehat{MSE}^{param} & if \ \left(\widehat{MSE}^{param}\right)^{-1} B^{-1} \sum_{b=1}^{B} u^{**(b,c)^2} < q \\ \widehat{MSE}^{db1} & otherwise \end{cases}, \tag{B11}$$

where $C = 1$ giving for the $b$th first level iteration only one value of $u^{**(b,c)}$, and the Authors' choice of $q$ value is 0.77.

Similarly, in [53, p. 3311] the formula of telescoping bootstrap MSE estimator (B8) is modified:

$$\widehat{MSE}^{dbTelEF}(\hat{\theta}) = \begin{cases} \widehat{MSE}^{param} & if \ \left(\widehat{MSE}^{param}\right)^{-1} B^{-1} \sum_{b=1}^{B} u^{**(b,c)^2} < q \\ \widehat{MSE}^{db-telesc} & otherwise \end{cases} \tag{B12}$$

where $C = 1$ and Authors assume that $q = 0.77$.

**Acknowledgements** The authors would like to express their gratitude to Professor Malay Ghosh from the University of Florida for discussions on partial results.

**Author Contributions** Conceptualization, proposal of the ex ante accuracy estimation methods, project and preparation of simulation studies, and the part of the paper and computations related to classic model-based methods were performed by Tomasz Żądło. Proposal of the plug-in predictor based on gradient boosting regression trees, data collection, and the part of the paper and computations concerning machine learning were performed by Adam Chwila. The first draft of the manuscript was written by both authors. Both authors read and approved the final manuscript.

**Funding** The work has been co-financed by the Minister of Science under the "Regional Initiative of Excellence" programme.

**Data Availability** Data used in the paper are freely available at https://bdl.stat.gov.pl.

## Declarations

**Conflict of interest** Authors have no Conflict of interest to declare that are relevant to the content of this article.

## References

1. Valliant R, Dorfman AH, Royall RM (2000) Finite population sampling and inference: a prediction approach. Wiley-Interscience, New York
2. Henderson CR (1950) Estimation of genetic parameters (Abstract). Ann Math Stat 21:309–310
3. Royall RM (1976) The linear least-squares prediction approach to two-stage sampling. J Am Stat Assoc 71:657–664. https://doi.org/10.1080/01621459.1976.10481542
4. Zakadlo T (2006) On accuracy of EBLUP under random regression coefficient model. Stat Trans 7:887–903
5. Pratesi M, Salvati N (2008) Small area estimation: the EBLUP estimator based on spatially correlated random area effects. Stat Methods Appl 17:113–141. https://doi.org/10.1007/s10260-007-0061-9
6. Molina I, Salvati N, Pratesi M (2009) Bootstrap for estimating the MSE of the spatial EBLUP. Comput Stat 24:441–458. https://doi.org/10.1007/s00180-008-0138-4
7. Chandra H, Salvati N, Chambers R, Tzavidis N (2012) Small area estimation under spatial nonstationarity. Comput Stat Data Anal 56:2875–2888. https://doi.org/10.1016/j.csda.2012.02.006
8. Krzciuk MK (2023) Small area estimation—model-based approach in economic research (University of Economics in Katowice, Katowice). https://doi.org/10.22367/uekat.9788378758860
9. Molina I, Rao JNK (2010) Small area estimation of poverty indicators. Can J Stat 38:369–385. https://doi.org/10.1002/cjs.10051
10. Jiang J, Nguyen T, Rao JS (2011) Best Predictive Small Area Estimation. J Am Stat Assoc 106:732–745. https://doi.org/10.1198/jasa.2011.tm10221
11. Sugasawa S, Kawakubo Y, Datta GS (2019) Observed best selective prediction in small area estimation. J Multivar Anal 173:383–392. https://doi.org/10.1016/j.jmva.2019.04.002
12. Chwila A, Zakadlo T (2022) On properties of empirical best predictors. Commun Stat Simul Comput 51:220–253. https://doi.org/10.1080/03610918.2019.1649422
13. Stachurski T (2021) Small area quantile estimation based on distribution function using linear mixed models. Econ Bus Rev 7:97–114. https://doi.org/10.18559/ebr.2021.2.7
14. Kontokosta CE, Hong B, Johnson NE, Starobin D (2018) Using machine learning and small area estimation to predict building-level municipal solid waste generation in cities. Comput Environ Urban Syst 70:151–162. https://doi.org/10.1016/j.compenvurbsys.2018.03.004
15. Chen J et al (2019) A comparison of linear regression, regularization, and machine learning algorithms to develop Europe-wide spatial models of fine particles and nitrogen dioxide. Environ Int 130:104934. https://doi.org/10.1016/j.envint.2019.104934
16. Jumin E et al (2020) Machine learning versus linear regression modelling approach for accurate ozone concentrations prediction. Eng Appl Comput Fluid Mech 14:713–725. https://doi.org/10.1080/19942060.2020.1758792
17. Zhang Y, Haghani A (2015) A gradient boosting method to improve travel time prediction. Transp Res Part C Emerg Technol 58:308–324. https://doi.org/10.1016/j.trc.2015.02.019
18. James G, Witten D, Hastie T, Tibshirani R (2013) An Introduction to Statistical Learning with Applications in R Vol. 103 of *Springer Texts in Statistics* (Springer, New York). https://doi.org/10.1007/978-1-4614-7138-7
19. Hastie T, Tibshirani R, Friedman J (2009) The elements of statistical learning. data mining, inference, and prediction. Springer Series in Statistics (Springer, New York). https://doi.org/10.1007/978-0-387-84858-7
20. Corral P, Molina I, Cojocaru A, Segovia S (2022) Guidelines to small area estimation for poverty mapping (World Bank). https://doi.org/10.1596/37728
21. Jean N et al (2016) Combining satellite imagery and machine learning to predict poverty. Science 353:790–794. https://doi.org/10.1126/science.aaf7894. (Publisher: American Association for the Advancement of Science)
22. Beck M, Dumpert F, Feuerhake J (2018) Machine learning in official statistics. arXiv preprint arXiv:1812.10422
23. Tzavidis N, Zhang L-C, Luna A, Schmid T, Rojas-Perilla N (2018) From start to finish: a framework for the production of small area official statistics. J R Stat Soc Ser A Stat Soc 181:927–979. https://doi.org/10.1111/rssa.12364
24. De Broe S et al (2020) Updating the paradigm of official statistics: New quality criteria for integrating new data and methods in official statistics. Working paper 02-20, Center for Big Data Statistics, Statistics Netherlands, The Hague. https://content.iospress.com/articles/statistical-journal-of-the-iaos/sji200711
25. Meertens QA, Diks CGH, Herik HJ (2022) Improving the output quality of official statistics based on machine learning algorithms. J Off Stat 38:485–508. https://doi.org/10.2478/jos-2022-0023
26. Breiman L (2001) Statistical modeling: the two cultures (with comments and a rejoinder by the author). Stat Sci 16:199–231. https://doi.org/10.1214/ss/1009213726
27. UNECE. Machine Learning for Official Statistics. Tech. Rep., The United Nations Economic Commission for Europe, Geneva (2021). https://unece.org/statistics/publications/machine-learning-official-statistics
28. Robinson C et al (2017) Machine learning approaches for estimating commercial building energy consumption. Appl Energy 208:889–904. https://doi.org/10.1016/j.apenergy.2017.09.060
29. Singleton A, Alexiou A, Savani R (2020) Mapping the geodemographics of digital inequality in Great Britain: An integration of machine learning into small area estimation. Comput Environ Urban Syst 82:101486. https://doi.org/10.1016/j.compenvurbsys.2020.101486
30. Krennmair P, Schmid T (2022) Flexible domain prediction using mixed effects random forests. J Roy Stat Soc: Ser C (Appl Stat) 71:1865–1894. https://doi.org/10.1111/rssc.12600

31. Chwila A (2023) The application of artificial intelligence models in commercial banks- opportunities and threats. Acta Universitatis Lodziensis. Folia Oeconomica 1:63–98. https://doi.org/10.18778/0208-6018.362.04
32. Vabalas A, Gowen E, Poliakoff E, Casson AJ (2019) Machine learning algorithm validation with a limited sample size. PLoS ONE 14:e0224365
33. AIML4OS | Eurostat CROS (2024). https://cros.ec.europa.eu/dashboard/aiml4os
34. Bojer CS, Meldgaard JP (2021) Kaggle forecasting competitions: An overlooked learning opportunity. Int J Forecast 37:587–603
35. Kontokosta CE, Hong B, Johnson NE, Starobin D (2018) Using machine learning and small area estimation to predict building-level municipal solid waste generation in cities. Comput Environ Urban Syst 70:151–162
36. Ledesma C, Garonita OL, Flores LJ, Tingzon I, Dalisay D (2020) Interpretable poverty mapping using social media data, satellite images, and geospatial information. arXiv preprint arXiv:2011.13563
37. Liu Y, Just A (2020) *SHAPforxgboost: SHAP Plots for 'XGBoost'* . https://github.com/liuyanguu/SHAPforxgboost/. R package version 0.1.0
38. Hajjem A, Bellavance F, Larocque D (2014) Mixed-effects random forest for clustered data. J Stat Comput Simul 84:1313–1328
39. Pellagatti M, Masci C, Ieva F, Paganoni AM (2021) Generalized mixed-effects random forest: a flexible approach to predict university student dropout. Stat Anal Data Min ASA Data Sci J 14:241–257
40. Dagdoug M, Goga C, Haziza D (2023) Model-assisted estimation through random forests in finite population sampling. J Am Stat Assoc 118:1234–1251. https://doi.org/10.1080/01621459.2021.1987250
41. Deville J-C, Särndal C-E (1992) Calibration estimators in survey sampling. J Am Stat Assoc 87:376–382. https://doi.org/10.1080/01621459.1992.10475217
42. Rao JNK, Molina I (2015) Small area estimation. Second edition, Wiley series in survey methodology (John Wiley & Sons, Inc, Hoboken, New Jersey)
43. Friedman JH (2001) Greedy function approximation: a gradient boosting machine. Ann Stat 29:1189–1232. https://www.jstor.org/stable/2699986. Publisher: Institute of Mathematical Statistics
44. González-Manteiga W, Lombardía M, Molina I, Morales D, Santamaría L (2007) Estimation of the mean squared error of predictors of small area linear parameters under a logistic mixed model. Comput Stat Data Anal 51:2720–2733. https://doi.org/10.1016/j.csda.2006.01.012
45. Flores-Agreda D, Cantoni E (2019) Bootstrap estimation of uncertainty in prediction for generalized linear mixed models. Comput Stat Data Anal 130:1–17. https://doi.org/10.1016/j.csda.2018.08.006
46. Żądło T (2013) in On parametric bootstrap and alternatives of MSE (ed. Vojáčková, H.) Proceedings of 31st International Conference Mathematical Methods in Economics 2013 1081–1086 (The College of Polytechnics Jihlava, Jihlava)
47. Wolny-Dominiak A, Zakadlo T (2022) On bootstrap estimators of some prediction accuracy measures of loss reserves in a non-life insurance company. Commun Stat Simul Comput 51:4225–4240. https://doi.org/10.1080/03610918.2020.1740263
48. González-Manteiga W, Lombardía MJ, Molina I, Morales D, Santamaría L (2008) Bootstrap mean squared error of a small-area EBLUP. J Stat Comput Simul 78:443–462. https://doi.org/10.1080/00949650601141811
49. Carpenter JR, Goldstein H, Rasbash J (2003) A novel bootstrap procedure for assessing the relationship between class size and achievement. J Roy Stat Soc: Ser C (Appl Stat) 52:431–443. https://doi.org/10.1111/1467-9876.00415
50. Chambers R, Chandra H (2013) A random effect block bootstrap for clustered data. J Comput Graph Stat 22:452–470. https://doi.org/10.1080/10618600.2012.681216
51. Thai H-T, Mentré F, Holford NH, Veyrat-Follet C, Comets E (2013) A comparison of bootstrap approaches for estimating uncertainty of parameters in linear mixed-effects models. Pharm Stat 12:129–140. https://doi.org/10.1002/pst.1561
52. Hall P, Maiti T (2006) On parametric bootstrap methods for small area prediction. J R Stat Soc Ser B (Stat Methodol) 68:221–238
53. Erciulescu AL, Fuller WA (2014) Parametric bootstrap procedures for small area prediction variance (ed.ASA) Proceedings of the Survey Research Methods Section 3307–3318 (The American Statistical Association, Boston, Massachusetts)
54. Pfeffermann D (2013) New important developments in small area estimation. Stat Sci 28:40–68. https://doi.org/10.1214/12-STS395
55. Norton EC, Karaca-Mandic P, Dowd B (2012) Interaction terms in nonlinear models. Health Serv Res 47:255–274
56. Gardini L, Radi D, Tramontana F (2023) Foreword of the special issue: nonlinear models and tools in economics, finance and social sciences. Commun Nonlinear Sci Numer Simul 119:107078
57. Buis ML (2010) Stata tip 87: Interpretation of interactions in nonlinear models. Stand Genom Sci 10:305–308
58. Örk Özel S, Çabuk HA (2023) Comparisons of the some estimators for the transcendental logarithmic (translog) model. Commun Stat-Simul Comput 52:4008–4022
59. Stadhouders N, Koolman X, van Dijk C, Jeurissen P, Adang E (2019) The marginal benefits of healthcare spending in the netherlands: Estimating cost-effectiveness thresholds using a translog production function. Health Econ 28:1331–1344
60. Mujuru NM, Obi A, Mishi S, Mdoda L (2022) Profit efficiency in family-owned crop farms in eastern cape province of south africa: a translog profit function approach. Agric Food Secur 11:20
61. Krzciuk M, Zadlo T (2014) On some tests of fixed effects for linear mixed models. Studia Ekonomiczne 189:49–57
62. Chen T, Guestrin C (2016) ACM (ed.) *XGBoost: a scalable tree boosting system.* (ed.ACM) *Proceedings of the 22nd ACM SIGKDD International Conference on Knowledge Discovery and Data Mining*, 785–794 (Association for Computing Machinery, New York). https://doi.org/10.1145/2939672.2939785
63. Bergstra J, Bengio Y (2012) Random search for hyper-parameter optimization. J Mach Learn Res 13:281–305. https://dl.acm.org/doi/10.5555/2188385.2188395
64. Jiang J (1996) Reml estimation: asymptotic behavior and related topics. *The Annals of Statistics* **24**, 255–286. https://www.jstor.org/stable/2242618
65. Bates D, Maechler M, Bolker B, Walker S, et al (2014) lme4: Linear mixed-effects models using eigen and s4. r package version 1.1-7. https://cran.r-project.org/web/packages/lme4/index.html
66. Davidson R, MacKinnon JG (2007) Improving the reliability of bootstrap tests with the fast double bootstrap. Comput Stat Data Anal 51:3259–3281. https://doi.org/10.1016/j.csda.2006.04.001

**Publisher's Note** Springer Nature remains neutral with regard to jurisdictional claims in published maps and institutional affiliations.